\documentclass[journal ]{new-aiaa}
\usepackage[utf8]{inputenc}
\usepackage{textcomp}

\usepackage{graphicx}
\usepackage{amsmath}
\usepackage[version=4]{mhchem}
\usepackage{siunitx}
\usepackage{longtable,tabularx}

\usepackage{bm}

\usepackage{amsmath,amssymb,amsfonts}

\usepackage{amsmath}

\usepackage{subcaption}

\usepackage{comment}

\usepackage{footnpag}  

\usepackage{caption}

\usepackage{graphicx}

\usepackage{float}

\usepackage{bm}

\usepackage{algpseudocode}

\usepackage{tikz}

\usepackage{caption,adjustbox,setspace}

\usepackage{float}

\usepackage{algorithm}

\algrenewcommand\algorithmicend{}%
\algtext*{EndIf}\algtext*{EndFor}\algtext*{EndWhile}%
\algtext*{EndProcedure}\algtext*{EndFunction}

\makeatletter
\algrenewtext{EndIf}{\algorithmicend\ \algorithmicif}
\algrenewtext{EndFor}{\algorithmicend\ \algorithmicfor}
\algrenewtext{EndWhile}{\algorithmicend\ \algorithmicwhile}

\makeatletter
\algrenewtext{EndIf}{\textbf{end if}}
\algrenewtext{EndFor}{\textbf{end for}}
\algrenewtext{EndWhile}{\textbf{end while}}
\makeatother

\makeatletter
\AtBeginEnvironment{algorithmic}{%
  \footnotesize
  \setlength{\itemsep}{0pt}%
  \setlength{\parskip}{0pt}%
  \setlength{\topsep}{0pt}%
  \setlength{\partopsep}{0pt}%
  \setlength{\abovedisplayskip}{2pt}%
  \setlength{\belowdisplayskip}{2pt}%
}
\makeatother

\usepackage{setspace}

\usetikzlibrary{arrows.meta, positioning, fit}

\usetikzlibrary{shapes.geometric, arrows.meta, positioning, fit, calc}

\usepackage{booktabs} 

\usepackage{multirow} 

\usepackage{tabularx}

\usepackage{booktabs}

\newcommand{\vect}[1]{\bm{#1}}

\newcommand{\mat}[1]{\bm{#1}}

\newcommand{\quat}[1]{\mathbf{#1}} 

\newcommand{\qmult}{\otimes} 
\usepackage{bm}

\usepackage{amsmath,amssymb,amsfonts}

\newcommand{\crs}[1]{\left[#1\right]_{\times}} 

\newcommand{\I}{\mathbf{I}}
\newcommand{\Zero}{\mathbf{0}}

\title{Compensating Star-Trackers Misalignments with Adaptive Multi-Model Estimation }

\author{Ridma Ganganath \footnote{Graduate Research Assistant, Department of Aerospace Engineering, Iowa State University, Ames, IA, and AIAA Member Grade (student)} and Simone Servadio\footnote{Assistant Professor, Department of Aerospace, Iowa State University, Ames, IA, and AIAA Member Grade (full) }}
\affil{Iowa State University, Ames, IA, 50014}

\author{David Daeyoung Lee\footnote{Assistant Professor, Department of Mechanical $\&$ Aerospace Engineering, California State University, Long Beach, CA, and AIAA Member }}
\affil{California State University, Long Beach, CA, 90840}

\begin{document}

\maketitle

\begin{abstract}
This paper presents an adaptive multi-model framework for jointly estimating spacecraft attitude and star-tracker misalignments in GPS-denied deep-space CubeSat missions. A Multiplicative Extended Kalman Filter (MEKF) estimates attitude, angular velocity, and gyro bias, while a Bayesian Multiple-Model Adaptive Estimation (MMAE) layer operates on a discrete grid of body-to-sensor misalignment hypotheses. In the single-misalignment case, the MEKF processes gyroscope measurements and TRIAD-based attitude observations, and the MMAE updates a three-dimensional grid over the misalignment vector. For a dual-misalignment configuration, the same MEKF dynamics are retained, and the MMAE bank is driven directly by stacked line-of-sight measurements from two star trackers, forming a six-dimensional grid over the two misalignment quaternions without augmenting the continuous-state dimension. A novel diversity metric, $\Psi$, is introduced to trigger adaptive refinement of the misalignment grid around a weighted-mean estimate, thereby preventing premature collapse of the model probabilities and concentrating computation in the most likely region of the parameter space. Monte Carlo simulations show arcsecond-level misalignment estimation and sub-degree attitude errors for both estimation problems, with estimation errors remaining well-bounded, proving robustness and consistency. These results indicate that the proposed MEKF--MMAE architecture enables accurate, autonomous, and computationally efficient in-flight calibration for resource-constrained spacecraft, and establishes dual star-tracker misalignment estimation as a practical option for deep-space CubeSat missions.
\end{abstract}

\section{Introduction}
\lettrine{A}{ccurate} and autonomous spacecraft attitude determination is a cornerstone of deep-space CubeSat missions, where limited ground support and the absence of GPS demand entirely onboard solutions. If increasing numbers of small satellite missions beyond Earth orbit exceed the capacity of NASA’s Deep Space Network, they must achieve high-fidelity position knowledge despite uncertain dynamics, sensor imperfections, and environmental disturbances \cite{mckeen2025attitude1,servadio2020nonlinear,doi:10.2514/1.G004355}. In this context, star trackers and gyroscopes are often the primary sensors for position and attitude estimation on resource-constrained platforms. However, practical challenges such as small misalignments between a star tracker’s mounting frame and the spacecraft body frame can introduce persistent errors in position estimation if not properly estimated and corrected. Ensuring robust attitude estimation via star trackers requires advanced alignment filtering architectures with attitude estimation that go beyond the classical single-model paradigm.

Traditional attitude filters like the Extended Kalman Filter (EKF) and its multiplicative variant (MEKF) have long been the industry standard due to their efficiency and ability to handle quaternion kinematics gracefully \cite{crassidis2007survey}. The MEKF, in particular, enforces the unit-quaternion constraint and avoids the singularities associated with Euler angles or other minimal attitude representations. While these filters perform well under nominal conditions, they typically assume a fixed sensor configuration and rely on a single model of the system. This single-model reliance can limit their robustness when confronted with unmodeled effects such as sensor misalignment, time-varying external disturbances, or non-Gaussian noise characteristics \cite{crassidis2004unscented}. In deep-space scenarios with minimal opportunities for ground recalibration, even a small star tracker misalignment can lead to a significant pointing error if the filter cannot adapt to it. Augmenting the state vector to estimate misalignment within a standard EKF/MEKF framework is possible but often poses convergence and observability issues, especially when the misalignment is constant and subtle.

To address these limitations, we propose a modular adaptive estimation framework that combines a 9-state MEKF with a Bayesian Multiple-Model Adaptive Estimation (MMAE) layer to jointly estimate the spacecraft’s attitude, gyroscope bias, angular velocity, and fixed star tracker misalignment. In the proposed architecture, the MEKF fuses gyroscope data and direction measurements, which are formulated as attitude observations via the TRIAD algorithm \cite{shuster1981three}. TRIAD provides a quaternion observation by fusing two non-collinear reference vectors (e.g., a sun vector and a star vector), offering a reliable attitude measurement source. This multi-model architecture avoids nonlinear state augmentation for misalignment estimation and supports parallel implementation, making it well-suited to CubeSat-class onboard processors. 

Our approach builds upon and extends several threads of prior research in adaptive filtering and spacecraft calibration. Earlier works have demonstrated the benefit of multiple-model methods in related contexts, such as adaptive calibration of star trackers and inertial sensors \cite{servadio2021differential}. For example, Lam et al. evaluated an MMAE scheme for improving on-orbit attitude determination accuracy by accounting for misalignment and other model uncertainties \cite{doi:10.2514/6.2007-6816}. Nebelecky and colleagues explored techniques for compensating model errors in attitude filters to enhance estimation fidelity under unknown disturbances \cite{6916281}. Crassidis et al. introduced a generalized MMAE framework that leverages residual autocorrelation to improve hypothesis discrimination in fault detection and calibration tasks \cite{4085937}. These studies underscore the value of model diversity and hypothesis testing in reliable spacecraft state estimation. Moreover, recent advances in uncertainty-aware filtering — such as the use of Koopman operator theory for nonlinear observer design under model uncertainty \cite{doi:10.2514/1.A35688} and particle filtering approaches that sample system dynamics from posterior distributions \cite{10643370} — motivate the inclusion of modular, robust inference architectures for complex systems. In contrast to prior methods, the framework presented here uniquely integrates a classical MEKF with a multi-hypothesis adaptation layer to handle sensor misalignments in real time, and introduces a quaternion-fusion step to maintain a high-confidence attitude solution drawn from multiple concurrent filters. 

In summary, this work provides a scalable and resilient attitude estimation solution for deep-space CubeSats facing sensor misalignment and other uncertainties. The proposed MEKF–MMAE architecture is capable of simultaneously estimating the spacecraft’s attitude, its angular velocity, and the true star tracker misalignment without requiring manual recalibration or state-vector augmentation. Monte Carlo simulations validate that the filter bank converges to the correct misalignment and consistently yields high-accuracy attitude knowledge, with errors remaining within $3\sigma$ uncertainty bounds. This adaptive, model-based approach thus enables high-fidelity autonomous navigation for resource-limited platforms, ensuring robust performance in GPS-denied environments where traditional single-model filters may falter.

\section{The Multiplicative Extended Kalman Filter with Star Tracker Measurement}
\label{partII}

We consider the spacecraft attitude estimation problem using a combination of two star-trackers and a gyroscope rigidly mounted on a CubeSat. The true attitude of the spacecraft is defined with respect to an inertial reference frame $\mathcal{I} = (x_I, y_I, z_I)$, fixed with respect to distant stars. The CubeSat body frame is denoted as $\mathcal{B} = (x_B, y_B, z_B)$ and is aligned with the spacecraft's structure. In this problem, the first star tracker, ST$_1$, observes a known inertial vector to Star 1 ($\vect{v}_1^{I}$) and provides a corresponding measurement, $\vect{v}_1^{B}$, in its own local sensor frame. Similarly, the second star tracker, ST$_2$, observes a known inertial vector to Star 2 ($\vect{v}_2^{I}$) and provides a measurement, $\vect{v}_2^{B}$, in its local frame. 
 
\begin{figure}[ht]
    \centering
    \includegraphics[width=0.5\linewidth]{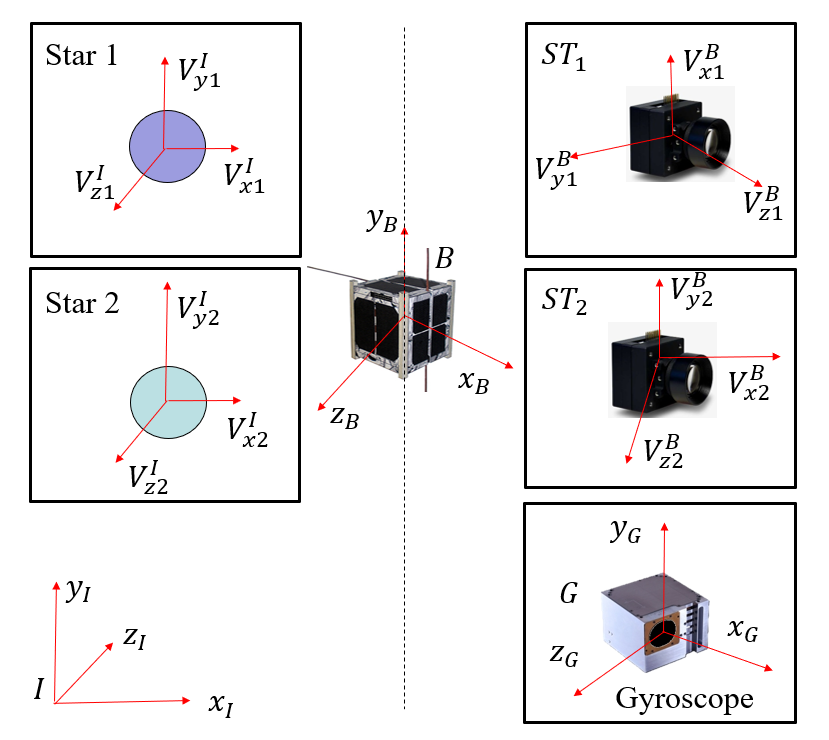} 
    \caption{Coordinate frames for the CubeSat (single misalignment case), including the inertial ($\mathcal{I}$), body ($\mathcal{B}$), gyroscope (G), and two star-tracker ($ST_1$, $ST_2$) reference frames.}
    \label{fig:framespic}
\end{figure}

Importantly, due to the nonlinearity and randomness, there will be many imperfections in the system.  In this work, the objective is to estimate the spacecraft's attitude $\mathbf{q}_{BI}(t)$, its angular velocity $\boldsymbol{\omega}(t)$, gyroscope bias $\mathbf{b}(t)$, and a constant star tracker misalignment vector $\boldsymbol{\mu}_R$. The estimation framework combines a 9-state Multiplicative Extended Kalman Filter (MEKF) to track the attitude, angular velocity, and gyroscope bias, with a Multiple-Model Adaptive Estimation (MMAE) layer to infer the misalignment $\boldsymbol{\mu}_R$ from a discrete grid of candidate hypotheses. The star tracker misalignment introduces a fixed rotational offset between the body frame and the sensor frame, affecting the measured directions of known inertial-frame vectors. Noisy measurements of these vectors, along with gyroscope readings, are used to update the filter and refine both attitude and misalignment estimates. Figure~\ref{fig:framespic} illustrates the relevant coordinate frames and sensor geometry involved in this joint estimation problem.

The MEKF serves as the foundational part of the architecture and is designed to estimate the spacecraft's attitude quaternion, angular velocity, and gyroscope bias in this problem. The filter operates on a multiplicative error state formulation to robustly handle the non-Euclidean nature of attitude quaternions. The true state of the system is defined as:
\begin{equation}
    \vect{x}_{\text{true}}(t) =
    \begin{bmatrix}
        \quat{q}(t) \\
        \vect{\omega}(t) \\
        \vect{b}(t)
    \end{bmatrix}
\end{equation}
where $\quat{q}(t) \in \mathbb{R}^4$ is the unit quaternion describing attitude (scalar-last convention), $\vect{\omega}(t) \in \mathbb{R}^3$ is the body-frame angular velocity, and $\vect{b}(t) \in \mathbb{R}^3$ represents the gyroscope bias.  The filter maintains a nominal state estimate, which can be described as:
\begin{equation}
\hat{\vect{x}}_{\text{nom}}(t) =
\begin{bmatrix}
\hat{\quat{q}}(t) \\
\hat{\vect{\omega}}(t) \\
\hat{\vect{b}}(t)
\end{bmatrix}
\end{equation}

The filter estimates a 9-dimensional error state, $\delta \vect{x}(t)$, which captures the deviation from the nominal estimate:
\begin{equation}\label{eq:3}
    \delta\vect{x}(t) =
    \begin{bmatrix}
        \delta\vect{\omega}(t) \\
        \delta\vect{b}(t) \\
        \delta\vect{\theta}(t)
    \end{bmatrix}
\end{equation}
where $\delta\vect{\omega}$ and $\delta\vect{b}$ are additive errors in the estimated angular velocity and bias, respectively. The attitude error is represented by a small-angle rotation vector, $\delta\vect{\theta} \in \mathbb{R}^3$, which relates the true and nominal quaternions via the multiplicative error quaternion described as:
\begin{equation}
    \delta\quat{q} \approx
    \begin{bmatrix}
        \frac{1}{2}\delta\vect{\theta} \\
        1
    \end{bmatrix}
\end{equation}
such that the corrected quaternion is evaluated via quaternion multiplication, can be defined as:
\begin{equation}
    \quat{q} = \delta\quat{q} \qmult \hat{\quat{q}}
\end{equation}
The uncertainty of the error state is captured by its covariance matrix, defined as:
\begin{equation}
    \mat{P}(t) = \mathbb{E}\left[\delta\vect{x}(t)\delta\vect{x}(t)^\top\right]
\end{equation}

\subsection{The System Dynamics}

The spacecraft’s true attitude is represented by a unit quaternion $\quat{q} \in \mathbb{R}^4$ (scalar-last convention) and the quaternion kinematics propagate as:
\begin{equation}
    \dot{\hat{\quat{q}}} = \frac{1}{2} \mat{\Omega}(\hat{\vect{\omega}}) \hat{\quat{q}} 
\end{equation}
where $\hat{\vect{\omega}} \in \mathbb{R}^3$ is the estimated body-frame angular velocity, and $\mat{\Omega}(\vect{\omega})$ is the quaternion multiplication matrix, which can be described as: 
\begin{equation}
    \mat{\Omega}(\vect{\omega}) =
    \begin{bmatrix}
        0 & -\omega_x & -\omega_y & -\omega_z \\
        \omega_x & 0 & \omega_z & -\omega_y \\
        \omega_y & -\omega_z & 0 & \omega_x \\
        \omega_z & \omega_y & -\omega_x & 0
    \end{bmatrix}
\end{equation}
In this formulation, the angular velocity evolves according to Euler’s rigid-body equations, with $M_c$ being the active external torques:
\begin{equation}
    \dot{\vect{\omega}} = \mat{J}^{-1} \left(M_c-\vect{\omega} \times (\mat{J}\vect{\omega}) \right)
\end{equation}
where $\mat{J}$ is the spacecraft's inertia matrix. The gyroscope bias $\vect{b} \in \mathbb{R}^3$ is modeled as constant:
\begin{equation}
    \dot{\vect{b}} = \mathbf{0}
\end{equation}

\subsection{The Prediction Step}
The MEKF is divided into the prediction step, when uncertainties are propagated under the dynamics, and a measurement update (or correction) step, where knowledge from the sensor is fused into the state uncertainties to obtain a more accurate estimate. 

The prediction step consists of two parallel operations. First, the nominal state $\hat{\vect{x}}_{\text{nom}}$ is propagated forward in time by numerically integrating the full nonlinear system dynamics. Simultaneously, the error-state covariance $\mat{P}$ is propagated using a linearized approximation based on the system Jacobians \cite{simon2006optimal}, which can be described as:
\begin{equation}
    \mat{P}_k^- = \mat{\Phi}_{k-1} \mat{P}_{k-1}^+ \mat{\Phi}_{k-1}^\top + \mat{Q} \Delta t
\end{equation}
where $\mat{P}_{k-1}^+$ denotes the a posteriori (updated) error covariance from the previous time step, and $\mat{P}_k^-$ represents the a priori (predicted) error covariance prior to incorporating the current measurement.
Here, $\mat{Q}$ denotes the continuous-time power spectral density, multiplied by $\Delta t$, which is the time increment between updates, to obtain the process noise covariance matrix under a Gaussian assumption.

We linearize about the nominal trajectory $(\hat{\boldsymbol\omega}(t),\hat {\boldsymbol q}(t))$ using a left–multiplicative attitude error with scalar-last quaternions and Modified Rodriguez Parameters (MRP) local coordinates. The state error is already introduced in Eq. \eqref{eq:3}, and between updates, we assume torque-free rigid-body motion with constant inertia $J$ as
\begin{equation}\label{eq:rb}
\dot{\boldsymbol\omega} \;=\; -\,\bm J^{-1}\,[\boldsymbol\omega]_\times (\bm J\boldsymbol\omega).
\end{equation}
Linearization at $\hat{\boldsymbol\omega}$ yields the standard rate–rate block
\begin{equation}\label{eq:Fww}
\bm F_{\boldsymbol\omega\boldsymbol\omega} \;=\; -\,\bm J^{-1}\!\Big([\hat{\boldsymbol\omega}]_\times \bm J - [\bm J\hat{\boldsymbol\omega}]_\times\Big),
\end{equation}
therefore, $\dot{\delta\boldsymbol\omega}=\bm F_{\boldsymbol\omega\boldsymbol\omega}\,\delta\boldsymbol\omega$. The bias is modeled as constant to first order, hence $\dot{\delta\boldsymbol b} = \boldsymbol 0$.
For the left–multiplicative MRP error, the small-angle kinematics gives as
\begin{equation}\label{eq:s_dot}
\dot{\delta\vect{\theta}} \;=\; -\,\tfrac14\,\delta\boldsymbol\omega,
\end{equation}
which shows that attitude error integrates the rate error at first order. Collecting the blocks in the state order $[\delta\boldsymbol\omega;\,\delta\boldsymbol b;\,\delta\boldsymbol\theta]$ gives the continuous-time error matrix as
\begin{equation}\label{eq:F_full}
\bm F \;=\;
\begin{bmatrix}
\bm F_{\boldsymbol\omega\boldsymbol\omega} & \Zero & \Zero\\[2pt]
\Zero & \Zero & \Zero\\[2pt]
-\tfrac14 I_3 & \Zero & \Zero
\end{bmatrix}.
\end{equation}
The full first-order kinematics include an attitude transport term $-[\hat{\boldsymbol\omega}]_\times\delta\vect{\theta}$, i.e.,
$\dot{\delta\vect{\theta}}=-(1/4)\,\delta\boldsymbol\omega - [\hat{\boldsymbol\omega}]_\times\delta\vect{\theta}$,
which would place $-[\hat{\boldsymbol\omega}]_\times$ in the $(\delta\vect{\theta},\delta\vect{\theta})$ block of $F$. For moderate $\|\hat{\boldsymbol\omega}\|$ and typical sampling periods, omitting this term has a negligible impact on estimation accuracy and yields a strictly block-lower-triangular $F$ that simplifies discretization and reduces the computational effort on the on-board computer \cite{doi:10.2514/1.G001217}.

Given a sampling interval $\Delta t$, the discrete-time error dynamics are described by the state transition matrix $\Phi = \exp(\bm F\Delta t)$. Because $F$ in Eq.~\eqref{eq:F_full} is block lower triangular in the state order $[\delta\boldsymbol{\omega};\,\delta\boldsymbol{b};\,\delta\vect{\theta}]$, the transition matrix $\Phi$ has the same block lower–triangular structure. In practice, this means that only the $3\times 3$ angular–rate block $\bm F_{\boldsymbol\omega\boldsymbol\omega}$ needs a full matrix exponential. The bias block is constant in time and therefore integrates to $I_3$, while the attitude error block follows directly from integrating Eq.~\eqref{eq:s_dot} over one sample:
\begin{equation}\label{eq:Phi}
\Phi \;=\;
\begin{bmatrix}
\exp(\bm F_{\boldsymbol\omega\boldsymbol\omega}\Delta t) & \Zero & \Zero\\[2pt]
\Zero & I_3 & \Zero\\[2pt]
-\tfrac14 I_3\,\Delta t & \Zero & I_3
\end{bmatrix}.
\end{equation}

This structure makes propagation explicit and computationally efficient.
The bias channel is identity at first order, and the attitude-error channel
advances by integrating the rate error. The continuous-time process noise is modeled with a block-diagonal covariance
\begin{equation}
\bm Q \;=\;
\begin{bmatrix}
\bm Q_\omega & \Zero & \Zero \\
\Zero & \bm Q_b & \Zero \\
\Zero & \Zero & \bm Q_s
\end{bmatrix},
\end{equation}
where $\bm Q_\omega$, $\bm Q_b$, and $\bm Q_s$ are $3\times 3$ diagonal matrices
tuned independently for the angular-rate, bias, and attitude-error
channels. The corresponding discrete-time process covariance is
\begin{equation}\label{eq:Qd_exact}
Q_d \;=\; \int_{0}^{\Delta t}\!\exp(\bm F\tau)\,\bm Q\,\exp(\bm F^\top\!\tau)\,d\tau.
\end{equation}
For sufficiently small~$\Delta t$, or when $Q$ is tuned conservatively
for robustness, the first-order approximation
\begin{equation}\label{eq:Qd_firstorder}
\bm Q_d \;\approx\; \bm Q\,\Delta t
\end{equation}
is adequate and preserves the intended filter tuning.

\subsection{The Measurement Model}
In this misalignment estimation problem, each measurement update, the spacecraft attitude is estimated using the TRIAD algorithm \cite{doi:10.2514/3.2555}, which computes the optimal rotation aligning two inertial-frame reference vectors with their corresponding measurements in the body frame. Let $\vect{v}_1^I$ and $\vect{v}_2^I$ be known reference vectors expressed in the inertial frame, and let $\vect{v}_1^B$ and $\vect{v}_2^B$ denote their noisy observations in the body frame. From these, two orthonormal TRIAD bases are constructed and can be described as:
\begin{align*}
    \vect{t}_1^I &= \frac{\vect{v}_1^I}{\|\vect{v}_1^I\|} & 
    \vect{t}_1^B &= \frac{\vect{v}_1^B}{\|\vect{v}_1^B\|} \\
    \vect{t}_2^I &= \frac{\vect{v}_1^I \times \vect{v}_2^I}{\|\vect{v}_1^I \times \vect{v}_2^I\|} &
    \vect{t}_2^B &= \frac{\vect{v}_1^B \times \vect{v}_2^B}{\|\vect{v}_1^B \times \vect{v}_2^B\|}\\
    \vect{t}_3^I &= \vect{t}_1^I \times \vect{t}_2^I &
    \vect{t}_3^B &= \vect{t}_1^B \times \vect{t}_2^B
\end{align*}
These vectors define the inertial-frame and body-frame TRIAD matrices and can be constructed as:
\begin{equation}
    \mat{T}_I = \begin{bmatrix} \vect{t}_1^I & \vect{t}_2^I & \vect{t}_3^I \end{bmatrix}, \quad
    \mat{T}_B = \begin{bmatrix} \vect{t}_1^B & \vect{t}_2^B & \vect{t}_3^B \end{bmatrix}
\end{equation}
The estimated direction cosine matrix (DCM) from inertial to body frame is computed as:
\begin{equation}
    \hat{\mat{C}}_{BI} = \mat{T}_B \mat{T}_I^\top
\end{equation}
which is then converted into a quaternion representation, $\quat{q}_{\text{meas}}$,, and serves as the TRIAD-based attitude measurement input for the filter update.

\subsection{The Update Step}
The measurement update step incorporates the TRIAD-derived quaternion and gyroscope measurements to correct the nominal state. The attitude residual quaternion is then computed as:
\begin{equation}
    \delta\quat{q}_{\text{res}} = \quat{q}_{\text{meas}} \qmult \hat{\quat{q}}^*
\end{equation}
where $\hat{\quat{q}}^*$ denotes the conjugate of the nominal quaternion estimate $\hat{\quat{q}}$. Assuming small attitude errors, this residual is converted into a Modified Rodrigues Parameter (MRP) vector:
\begin{equation}
    \delta\vect{\theta}_{\text{res}} = \text{q2mrp}(\delta\quat{q}_{\text{res}})
\end{equation}
where the $\text{q2mrp}(\cdot)$ function indicates the classic change in attitude representation from quaternions to MRP. The angular velocity residual is defined as the difference between the measured angular velocity and the predicted value\cite{KOTTATH201688}:
\begin{equation}
\vect{\omega}_{\text{res}} = \vect{\omega}_{\text{meas}} - (\hat{\vect{\omega}} + \hat{\vect{b}})
\end{equation}
where $\hat{\vect{b}}$ accounts for the estimated gyroscope bias. The residual of the composite measurement becomes:
\begin{equation}
    \vect{y} =
    \begin{bmatrix}
        \delta\vect{\theta}_{\text{res}} \\
        \vect{\omega}_{\text{res}}
    \end{bmatrix}, \quad
    \mat{H} =
    \begin{bmatrix}
        \mathbf{0} & \mathbf{0} & \mat{I} \\
        \mat{I} & \mat{I} & \mathbf{0}
    \end{bmatrix}
\end{equation}
so that the measurements linearly inform the state, being $\mat{H}$ the linearized measurement Jacobian.  The innovation covariance $\mat{S}$ and Kalman gain $\mat{K}$ are computed as:
\begin{align}
    \mat{S} &= \mat{H} \mat{P}^- \mat{H}^\top + \mat{R} \\
    \mat{K} &= \mat{P}^- \mat{H}^\top \mat{S}^{-1} 
\end{align}
where $\mat{P}^-$ is the prior error covariance, and $\mat{R}$ is the measurement noise covariance. This leads to the state and covariance update equations
\begin{align}
    \delta\vect{x}^+ &= \mat{K} \vect{y} \\
    \mat{P}^+ &= (\mat{I} - \mat{K} \mat{H}) \mat{P}^- (\mat{I} - \mat{K} \mat{H})^\top + \mat{K} \mat{R} \mat{K}^\top
\end{align}
The attitude estimate is now corrected using the small-angle approximation, where the attitude error component $\delta\vect{\theta}^+$ is extracted from $\delta\vect{x}^+$. This vector is converted into a quaternion correction and described as:
\begin{align}
    \delta\quat{q}^+ &= \text{mrp2q}(\delta\vect{\theta}^+) \\
    \hat{\quat{q}}^+ &= \delta\quat{q}^+ \qmult \hat{\quat{q}}^-, \quad \hat{\quat{q}}^+ \leftarrow \frac{\hat{\quat{q}}^+}{\|\hat{\quat{q}}^+\|}
\end{align}
while the nominal angular velocity, $\hat{\vect{\omega}}$, and bias, $\hat{\vect{b}}$, are updated with their corresponding components. Finally, the error state is reset to zero, which completes the measurement update cycle and ensures the quaternions are never directly added up, as they represent rotations.

\subsection{The Measurement Noise Models}
To assess the MEKF's sensitivity to sensor noise, we consider two prevalent noise models applied to the body-frame inputs of the TRIAD algorithm: an additive model and a multiplicative model.

The additive model perturbs each rotated inertial vector with zero-mean Gaussian noise and can be described as:
\begin{align}
    \vect{v}_i^B &= \mat{C}_{BI} \vect{v}_i^I + \vect{\eta}_i 
\end{align}
where $\mat{C}_{BI}$ is the true direction cosine matrix (DCM) from inertial to body frame, $\vect{v}_i^I$ are known inertial vectors ($i = 1,2$), and $\vect{\eta}_i \sim \mathcal{N}(\vect{0}, \sigma_\eta^2 \mat{I}_3)$ is zero-mean Gaussian vector noise. The vector is then normalized to respect unity constraints.

In contrast, the multiplicative model introduces uncertainty through small random rotations, and can be stated as:
\begin{align}
    \vect{v}_i^B &= \mat{C}_{\eta_i} \mat{C}_{BI} \vect{v}_i^I
\end{align}
where $\mat{C}_{\eta_i} = \text{mrp2dcm} (\delta\vect{\eta}_i)$ is a small random rotation matrix generated from zero-mean attitude noise, $\delta\vect{\eta}_i \sim \mathcal{N}(\vect{0}, \sigma_\eta^2 \mat{I}_3)$, with $\text{mrp2dcm}(\cdot)$ the function indicates the classic change in attitude representation from MRP to the direct cosine matrix. Multiplicative measurement noise is more realistic when compared to its additive counterpart, as rotations are stacked in series via multiplication rather than addition \cite{doi:10.2514/1.G003221}.

\section{The Robust Multiple-Model Adaptive Estimation (MMAE) Framework for Joint Attitude and Star Tracker Misalignment Estimation }

This approach augments the classical Multiplicative Extended Kalman Filter (MEKF) by estimating the star tracker misalignment vector, $\boldsymbol{\mu}_R$, using a grid-based Multiple Model Adaptive Estimation (MMAE) framework. During installation, star trackers might have an offset misalignment of the hardware with respect to the assumed direction in the software. The proposed MMAE accounts for such deviations and proposes a model-based estimation. Attitude is estimated using the TRIAD method, and the optimal quaternion is computed via the Markley quaternion fusion \cite{markley2007averaging}, later derived. Each model in the MMAE bank corresponds to a hypothesis of $\boldsymbol{\mu}_R$.

Let $\bm{y}_k \in \mathbb{R}^m$ denote the measurement at time step $k$, such as a TRIAD-based attitude quaternion converted to a Modified Rodrigues Parameters vector. Assume a bank of $N$ competing models $\mathcal{M}_j$, each corresponding to a distinct hypothesis of the star tracker misalignment vector $\bm{\mu}_R^{(j)} \in \mathbb{R}^3$. Let $\hat{\bm{y}}_k^{(j)}$ denote the predicted measurement under model $\mathcal{M}_j$, such that $\bm{r}_j = \bm{y}_k - \hat{\bm{y}}_k^{(j)}$ is the residual under model $j$, $\bm{R} \in \mathbb{R}^{m \times m}$ the measurement noise covariance matrix, and $w_j^{(k-1)}$ the prior probability (weight) of model $\mathcal{M}_j$ at time $k-1$.

Once an observation is received and processed, $\bm{y}_k$, the aim is to compute the posterior model probability via a Bayesian approach, defined as:
\begin{equation}
P(\mathcal{M}_j \mid \bm{y}_k) = \frac{ P(\bm{y}_k \mid \mathcal{M}_j) P(\mathcal{M}_j) }{ P(\bm{y}_k) }
\end{equation}
Using the model prior weight, which is its probability after prediction, $P(\mathcal{M}_j) = w_j^{(k-1)}$, and marginal likelihood $P(\bm{y}_k) = \sum_{h=1}^N P(\bm{y}_k \mid \mathcal{M}_h) w_h^{(k-1)}$, we get the weight update equation as:
\begin{equation}
w_j^{(k)} =
\frac{
w_j^{(k-1)} \cdot P(\bm{y}_k \mid \mathcal{M}_j)
}{
\sum\limits_{h=1}^{N} w_h^{(k-1)} \cdot P(\bm{y}_k \mid \mathcal{M}_h)
}
\end{equation}
Assuming additive zero-mean Gaussian measurement noise in the measurement model, regardless of how the true measurement is affected by the stochastic variable, the likelihood of measurement $\bm{y}_k$ under model $\mathcal{M}_j$ is:
\begin{equation}
    P(\bm{y}_k \mid \mathcal{M}_j) =
    \frac{1}{(2\pi)^{m/2} \sqrt{\det \bm{R}}} \exp \left( -\frac{1}{2} \bm{r}_j^\top \bm{R}^{-1} \bm{r}_j \right)
\end{equation}
Substituting into the Bayesian weight update then gives:
\begin{equation}
    w_j^{(k)} =
    \frac{
    w_j^{(k-1)} \cdot \exp \left( -\frac{1}{2} \bm{r}_j^\top \bm{R}^{-1} \bm{r}_j \right)
    }{
    \sum\limits_{h=1}^{N} w_h^{(k-1)} \cdot \exp \left( -\frac{1}{2} \bm{r}_h^\top \bm{R}^{-1} \bm{r}_h \right)
    }
\end{equation}
which provides the Maximum Likelihood-based Bayesian model probability update.

\subsection{Implementation for Attitude Residuals}
The residual for each model is the rotational error between the TRIAD-based measurement quaternion, $\bm{q}_{\text{meas}}$, and the quaternion predicted by that model, $\bm{q}_j$. This error is expressed as a three-dimensional small-angle vector in the MRP domain, and can be described as
\begin{equation}
\vect{s}_{\text{res}}^{(j)} = \text{q2MRP}\left( \bm{q}_{\text{meas}} \otimes \hat{\bm{q}}_j^{*} \right)
\end{equation}

To maintain numerical stability, the weight update is performed in two separate steps. First, an un-normalized weight is computed by multiplying the prior weight by the likelihood of the residual, which is assumed to be Gaussian, stated as:
\begin{equation}
\tilde{w}_j^{(k)} = w_j^{(k-1)} \cdot \exp\left(-\frac{1}{2} \bm{r}_j^\top \bm{R}^{-1} \bm{r}_j \right)
\end{equation}
This intermediate value incorporates the new measurement information. The final weight is then found by normalizing across all models to ensure the weights sum to one:
\begin{equation}
w_j^{(k)} =
\frac{
\tilde{w}_j^{(k)}
}{
\sum\limits_{h=1}^{N} \tilde{w}_h^{(k)}
}
\end{equation}
This two-step process provides a robust, recursive update rule for the model probabilities.

\subsection{Pruning}

Over time, many hypotheses become irrelevant (i.e., have low weight). To improve computational efficiency, a pruning threshold, $w_\text{prune}$, is introduced to eliminate them. First, the set of all valid hypotheses is identified as those whose weights exceed this threshold as:
\begin{equation}
\mathcal{J}_\text{valid}^{(k)} = \left\{ j \mid w_j(k) > w_\text{prune} \right\}
\end{equation}
This step effectively creates a list of all models that remain consistent with the measurement data. The filter bank is then reduced, retaining only these statistically significant hypotheses:
\begin{equation}
\left\{ \bm{\mu}_R^{(j)} \right\}_{j=1}^N \rightarrow \left\{ \bm{\mu}_R^{(j)} \right\}_{j \in \mathcal{J}_\text{valid}^{(k)}}
\end{equation}
Finally, the weights of the all surviving models are renormalized to ensure their sum remains one, preserving a valid probability distribution, can be desceibed as:

\begin{equation}
w_j(k) \leftarrow \frac{ w_j(k) }{ \sum\limits_{h \in \mathcal{J}_\text{valid}^{(k)}} w_h(k) }, \quad \forall j \in \mathcal{J}_\text{valid}^{(k)}
\end{equation}
This complete pruning process reduces computational complexity in subsequent updates by focusing the limited computational resources only on the most probable models.

\subsection{Grid Branching and Refinement Strategy}
Estimating the sensor misalignment presents a conflict between precision and computational cost, especially for CubeSat deep space navigation. A globally high-resolution hypothesis grid is computationally infeasible, while a coarse grid is inaccurate. Adaptive refinement resolves this by starting coarse and iteratively ``zooming in" on the most probable solution. This strategy achieves a high-precision estimate by focusing computational resources only where it is needed, making the problem tractable and efficient.

The trigger for this refinement is a crucial design choice. In this work, we investigate and compare three distinct triggering mechanisms to determine the most robust strategy.

The first approach is the classical trigger centered at the dominant model. This is the most common and classical selection, which is based on implementing a refinement when a single model becomes dominant, identified by its weight exceeding a predefined threshold. Let \( w_j^{(k)} \) denote the weight of the \( j \)-th hypothesis \( \boldsymbol{\mu}_R^{(j)} \) at time step \( k \). The first trigger condition is met when the maximum model weight exceeds a predefined branching threshold \( w_{\text{branch}} \);
\begin{equation}
\max_j w_j^{(k)} > w_{\text{branch}}
\label{eq:branch_threshold}
\end{equation}
Once the condition in Eq.~\eqref{eq:branch_threshold} is satisfied, the Maximum A Posteriori (MAP) index \( j^* \) and the corresponding misalignment hypothesis are
\begin{equation}
  j^* = \arg\max_j w_j^{(k)}, \qquad
  \hat{\boldsymbol{\mu}}_R^{\text{MAP}} = \boldsymbol{\mu}_R^{(j^*)}.
  \label{eq:map_estimate}
\end{equation}
At that instant, a new locally refined hypothesis grid is constructed, centered at \( \hat{\boldsymbol{\mu}}_R^{\text{MAP}} \) and with a
smaller spacing between neighboring models. Each model in this refined grid is initialized with the MEKF state associated with the MAP hypothesis in the center, and the corresponding model probabilities are reset to a uniform distribution.

The second approach consists of the creation of a performance index, the hypothesis diversity \(\Psi(t)\) trigger. Indeed, we propose a more robust dual-trigger strategy. In addition to the maximum weight criterion, we monitor the overall hypothesis diversity using the $\Psi$ metric, which comes from an analogy to the effective number of particles in particle filters \cite{10643370}. To quantify model diversity over time, we define the weight uniformity metric
\begin{equation}
    \Psi(t) \;=\; 100 \cdot \frac{A_t}{N},
    \label{eq:psi}
\end{equation}
with
\begin{equation}
    A_t \;=\; \left( \sum_{i=1}^N w_i^2(t) \right)^{-1},
    \label{eq:psi_At}
\end{equation}
where $N$ is the number of models in the current hypothesis set and $w_i(t)$ is the normalized weight of model $i$ at time $t$, so that $\sum_{i=1}^N w_i(t) = 1$.
 The term \(A_t\) represents the inverse of the weight concentration (also known as the effective number of models). A value of \(\Psi(t) = 100\%\) implies a perfectly uniform model distribution, while low $\Psi$ value indicates that the filter has successfully localized the solution, even if no single model is dominant.  When the refinement is triggered by a low $\Psi$ value, the new grid is centered on the hypothesis with the Maximum A Posteriori (MAP) probability at that instant. This prevents the filter from getting stuck in a state of local consensus where no single model is confident enough to trigger a refinement on its own, but a direction of likely vectors is highlighted. 

The third approach builds on the second. This final strategy utilizes the diversity-based trigger and enhances the grid recentering mechanism. While the trigger condition remains the same, $\Psi(t) = 10\%$ in the applications, the new refined grid is centered on the weighted mean of the hypothesis set rather than the MAP estimate, which is described in Equation \eqref{eq:mean_estimate}, described as:
\begin{equation}
\hat{\boldsymbol{\mu}}_R^{\text{Mean}} = \sum_{j=1}^{N} w_j^{(k)} \boldsymbol{\mu}_R^{(j)}
\label{eq:mean_estimate}
\end{equation}
Therefore, a new grid with finer space among models is centered at the models' weighted mean, with the weight redistributed among all the new models to continue improving accuracy. The process repeats with increasingly finer grids until convergence is achieved.

\subsection{Optimal Attitude Fusion via Quaternion Averaging}
At each time step, the MMAE filter produces a bank of \( N \) distinct attitude quaternions, \( \hat{\quat{q}}^{(j)} \), each with an associated posterior probability, \( w^{(j)} \). The final step is to fuse this distributed information into a single, optimal attitude estimate that best represents the entire set of hypotheses. For this, we employ the statistically optimal quaternion averaging method developed by Markley \cite{averageq}, as the quaternion mean cannot be evaluated simply by addition due to their unity constraint. 

The method seeks the mean quaternion, \( \hat{\quat{q}} \), that minimizes the weighted sum of the squared chordal distances to each hypothesis quaternion in the bank, which can be explained as
\begin{equation}
\hat{\quat{q}} = \arg\min_{\quat{q} \in \mathbb{S}^3} \sum_{j=1}^{N} w^{(j)} \left\| \quat{q} - \hat{\quat{q}}^{(j)} \right\|^2
\label{eq:min_chordal}
\end{equation}
The literature demonstrated that this optimization problem can be elegantly solved by constructing a \( 4 \times 4 \) symmetric matrix, \( \mat{M} \), which aggregates the weighted outer products of all attitude hypotheses, which can be described as:
\begin{equation}
\mat{M} = \sum_{j=1}^{N} w^{(j)} \hat{\quat{q}}^{(j)} \left( \hat{\quat{q}}^{(j)} \right)^T
\label{eq:markley_matrix}
\end{equation}
Then the optimal fused quaternion, \( \hat{\quat{q}} \), that satisfies the minimization criterion in Eq. \eqref{eq:min_chordal} is the eigenvector of \( \mat{M} \) corresponding to its maximum eigenvalue, given as:
\begin{equation}
\hat{\quat{q}} = \mathrm{eigvec}_{\max}(\mat{M})
\label{eq:markley_fused_q}
\end{equation}
To ensure the result remains a valid unit quaternion on the hypersphere \( \mathbb{S}^3 \), a final normalization is performed to enhance the filter's robustness. Furthermore, to address the inherent sign ambiguity of the quaternion representation (\( \quat{q} \equiv -\quat{q} \)), the resulting quaternion is aligned to maintain consistency with the previous time step’s estimate. This fusion process, combined with the adaptive grid refinement, forms the core of a robust MEKF-MMAE framework capable of accurate joint estimation.

\subsection{System Architecture}

\begin{figure}[H]
  \centering
  \includegraphics[width=0.99\linewidth]{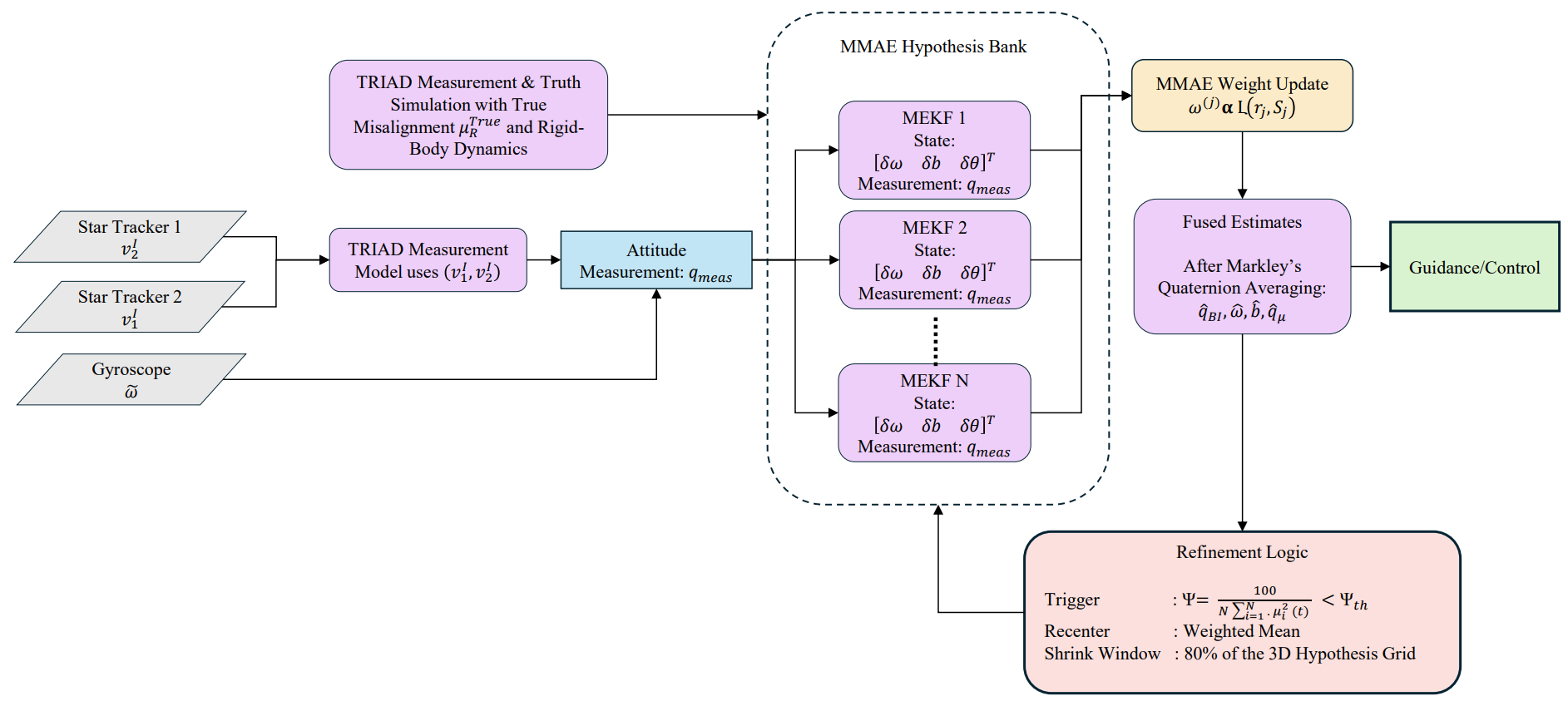}
  \caption{Single-misalignment MEKF–MMAE architecture. Likelihoods update model weights $w^{(j)}$, which govern grid refinement and Markley quaternion fusion.}
  \label{fig:mmae_single}
\end{figure}

Figure~\ref{fig:mmae_single} summarizes the misalignment estimation MEKF--MMAE architecture introduced in Sec.~\ref{partII}, while Algorithm~\ref{alg:mmae_mekf} outlines one update cycle. Star-tracker and gyro measurements are processed by a bank of MEKFs, each conditioned on a different misalignment hypothesis, and the corresponding innovations are converted to likelihoods that update the model weights. The normalized weights drive both the adaptive grid–refinement logic and the quaternion averaging step, yielding a single fused attitude, rate, bias estimate, and single-misalignment estimate for guidance and control.

\setlength{\textfloatsep}{8pt}
\setlength{\floatsep}{5pt}
\setlength{\intextsep}{6pt}

\begin{algorithm}[H]
\caption{The MMAE Algorithm for Attitude and Misalignment Estimation}
\label{alg:mmae_mekf}
\small 
\begin{algorithmic}[1]
\begin{spacing}{0.95} 
\State \textbf{Initialize:} Nominal state $(\hat{\quat{q}}_0, \hat{\boldsymbol{\omega}}_0, \hat{\boldsymbol{b}}_0)$ and covariance $\mat{P}_0$
\State \textbf{Hypotheses:} Generate a small-angle misalignment grid $\{\boldsymbol{\mu}_R^{(j)}\}_{j=1}^N$ and assign uniform weights $w_j = 1/N$
\State \textbf{Filter Bank:} Initialize $N$ MEKF instances, each corresponding to a hypothesis $\boldsymbol{\mu}_R^{(j)}$
\For{each timestep $k$}
    \State Propagate truth state and generate measurements $(\quat{q}_{\text{meas}}, \boldsymbol{\omega}_{\text{meas}})$
    \For{each model $j$}
        \State Predict MEKF state $(\hat{\quat{q}}^{(j)}, \hat{\boldsymbol{\omega}}^{(j)}, \hat{\boldsymbol{b}}^{(j)})$ and form expected measurement $\hat{\quat{q}}_{\text{exp}}^{(j)} = \quat{q}_{\mu}^{(j)} \otimes \hat{\quat{q}}^{(j)}$
        \State Compute residual $\mathbf{y}_k^{(j)} = [\text{MRP}(\quat{q}_{\text{meas}} \otimes \hat{\quat{q}}_{\text{exp}}^{(j),-1}); \ \boldsymbol{\omega}_{\text{meas}} - (\hat{\boldsymbol{\omega}}^{(j)} + \hat{\boldsymbol{b}}^{(j)})]$
        \State Update MEKF state using residual $\mathbf{y}_k^{(j)}$
        \State Update weight: $w_j \gets w_j \cdot L(\mathbf{y}_k^{(j)})$
    \EndFor
    \State Normalize weights: $w_j \gets w_j / \sum_{l=1}^N w_l$
    \State Compute diversity metric: $\Psi_k = \frac{100}{N \sum_j w_j^2}$
    \If{$\Psi_k < \Psi_{\text{th}}$ and refinements remain}
        \State Compute weighted mean: $\hat{\boldsymbol{\mu}}_R = \sum_j w_j \boldsymbol{\mu}_R^{(j)}$ and generate refined hypothesis grid centered at $\hat{\boldsymbol{\mu}}_R$
         
        \State Reset weights to uniform  $w_j = 1/N$
    \Else
        \State  Prune low-weight models and re-normalize weights
    \EndIf
    \State Fuse attitude via optimal quaternion averaging
    \State Store state estimates and misalignment error
\EndFor
\end{spacing}
\end{algorithmic}
\end{algorithm}

\section{Numerical Simulations: Single Misalignment Scenario}
Our entire problem in this work was validated and simulated using non-linear rigid-body dynamics, as described in the problem formulation. The simulation initializes the spacecraft in a high-rotation condition, with a constant angular velocity vector of $\boldsymbol{\omega} = [3.0, 4.4, -5.0]^T$ deg/s, emulating a steady rotation phase to avoid unobservable scenarios. To replicate realistic post-maneuver behavior, a damping torque is applied after a predefined time duration, as later reported. This braking torque is modeled as $\boldsymbol{\tau}_\text{damp} = -D \boldsymbol{\omega}$, where $D = 0.6$ is a scalar damping coefficient. The full angular velocity dynamics thus become:
\begin{equation}
\dot{\boldsymbol{\omega}} = \mathbf{J}^{-1} \left( \boldsymbol{M}_c - \boldsymbol{\omega} \times (\mathbf{J} \boldsymbol{\omega}) - D \boldsymbol{\omega} \right)
\quad
\label{manu}
\end{equation}
causing the spacecraft's spin rate to decay gradually toward zero. 

In this misalignment estimation problem, a Monte Carlo analysis with 100 runs was conducted, each lasting 5000 seconds with a 0.5 seconds time step, for a spacecraft with an inertia matrix of 
$\bm{J} = \mathrm{diag}(100,\ 60,\ 50)\ \mathrm{kg} \cdot \mathrm{m}^2$. Moreover, the breaking torque has been introduced after 4100 seconds.  The filter's tuning parameters, including the initial covariance ($\bm{P}_0$), process noise ($\bm{Q}$), and measurement noise ($\bm{R}$), are detailed in Table~\ref{tab:covariance_matrices}. 
For the adaptive estimation, the process began with a $7 \times 7 \times 7$ three-dimensional hypothesis grid with an initial step size of $\Delta_\mu \approx 0.167^\circ$ on each axis. A hypothesis diversity $\Psi$ threshold of 10\% to trigger the refinement has been chosen.

We evaluate the MEKF–MMAE framework for a single fixed star-tracker misalignment. A bank of MEKFs is executed in parallel, one per hypothesis $\bm{\mu}_R^{(j)}$. All filters ingest the same TRIAD attitude observation and gyroscope rate. For each hypothesis, the stacked innovation (TRIAD small-angle residual + gyro residual) yields a likelihood that updates the model weight $w^{(j)}$. These weights drive (i) the pruning/refinement of the hypothesis grid, and (ii) the Markley weighted quaternion fusion for the bank attitude. 

\begin{table}[h]
\centering
\caption{Covariance and Noise Matrix Values}
\label{tab:covariance_matrices}
\begin{tabular}{@{} l l c @{}} 
\toprule
\textbf{Matrix} & \textbf{Component} & \textbf{Value Used} \\
\midrule
\multirow{3}{*}{\bm{$P_0$}} & Angular Velocity & $(0.01)^2 \ \mathrm{rad}^2/\mathrm{s}^2$ \\
 & Gyroscope Bias & $(0.001)^2 \ \mathrm{rad}^2/\mathrm{s}^2$ \\
 & Attitude Error (MRP) & $(1.0)^2 \ \mathrm{rad}^2$ \\
\midrule
\multirow{3}{*}{\bm{$Q$}} & Angular Velocity & $(1 \times 10^{-6})^2 \ \mathrm{rad}^2/\mathrm{s}^4$ \\
 & Gyroscope Bias Drift & $(5 \times 10^{-8})^2 \ \mathrm{rad}^2/\mathrm{s}^4$ \\
 & Attitude Drift & $(5 \times 10^{-7})^2 \ \mathrm{rad}^2$ \\
\midrule
\multirow{2}{*}{\bm{$R$}} & Star Tracker (Attitude) & $(8.73 \times 10^{-4})^2 \ \mathrm{rad}^2$ \\
 & Gyroscope (Rate) & $(0.0005)^2 \ \mathrm{rad}^2/\mathrm{s}^2$ \\
\bottomrule
\end{tabular}
\end{table}

\subsection{Additive Vs. Multiplicative Results and Analysis}

\begin{figure}[H]
    \centering
    \includegraphics[width=0.55\textwidth]{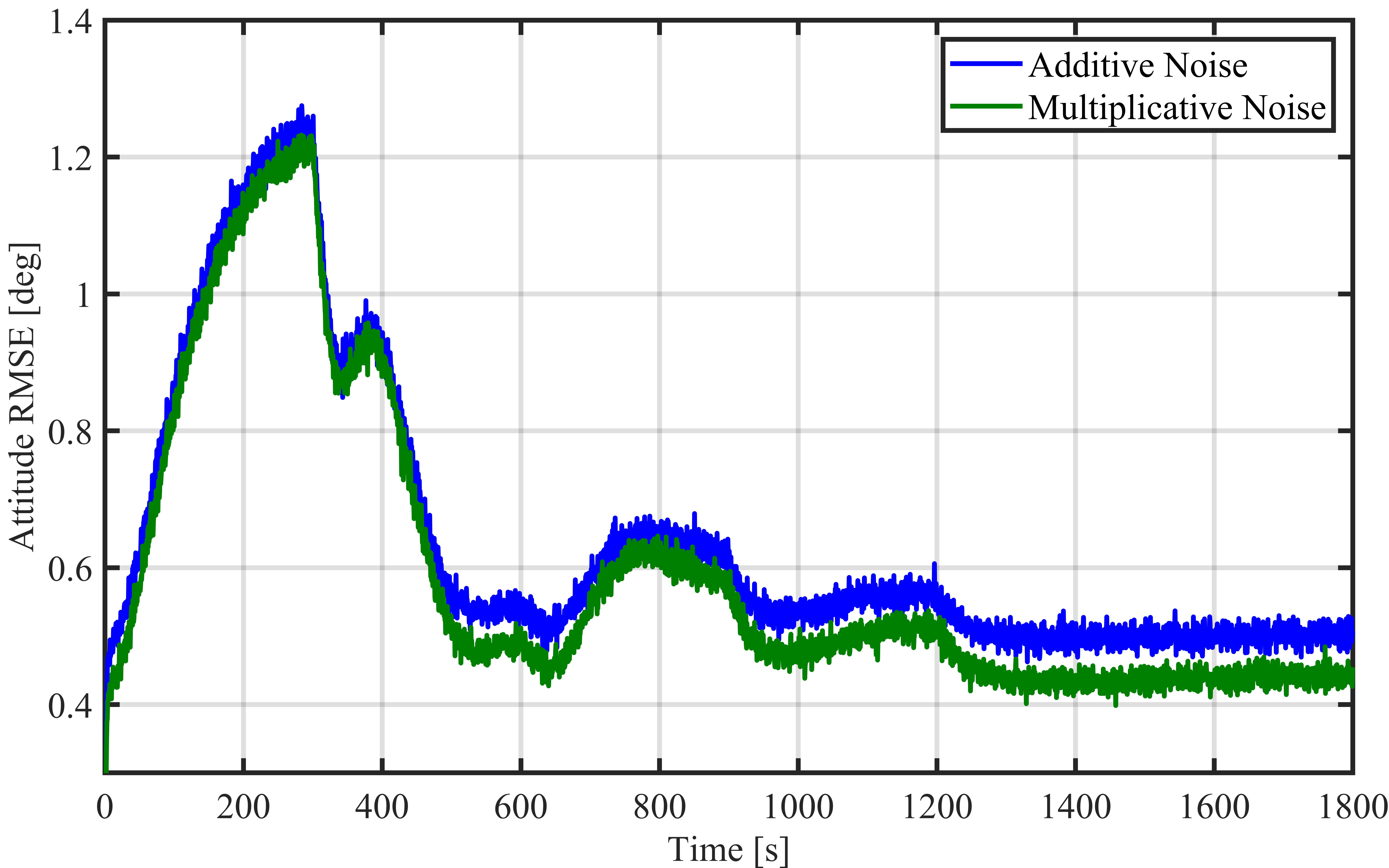}
    \caption{Attitude RMSE comparison: additive vs. multiplicative TRIAD noise.}
    \label{fig:additive_vs_multiplicative}
\end{figure}

The MEKF's performance was evaluated under both additive and multiplicative TRIAD noise models to analyze the physical difference in estimation, considering the provided level of noise from the star tracker datasheet. The results of this comparison, averaged over 100 Monte Carlo runs, are presented in Figure~\ref{fig:additive_vs_multiplicative} and quantified in
Table~\ref{tab:noise_comparison}, where the CubeSat performs maneuvers to test different angular velocities. At each time step, the attitude error for run $i$ is taken as the principal rotation angle of the quaternion that maps the estimated attitude to the true attitude. The plotted RMSE is
\begin{equation}
\label{eq:att_rmse}
\theta_{\mathrm{RMSE}}(t_k)
\;=\;
\sqrt{\frac{1}{N_{\mathrm{MC}}}
\sum_{i=1}^{N_{\mathrm{MC}}} \theta_i^2(t_k)},
\end{equation}
expressed in degrees.

\begin{table}[H]
    \centering
    \caption{Final attitude error comparison: additive vs. multiplicative TRIAD noise.}
    \label{tab:noise_comparison}
    \begin{tabular}{@{}lcc@{}}
        \toprule
        \textbf{Metric} & \textbf{Additive} & \textbf{Multiplicative} \\
        \midrule
        Mean Final Error (deg) & 0.4640 & 0.3887 \\
        Std Final Error (deg)  & 0.1915 & 0.1663 \\
        Max Final Error (deg)  & 1.1207 & 1.1217 \\
        \bottomrule
    \end{tabular}
\end{table}

To ensure the filter was tested in a realistic scenario, a series of maneuvers was simulated in this specific MEKF example, with step changes in the spacecraft's true angular velocity occurring at 5, 10, 15, and 20 minutes into the simulation. Under these dynamic conditions, the multiplicative model has a lower influence on attitude estimation accuracy compared to the additive model at the same power spectral density level. Specifically, it achieves a 16\% lower mean final error (0.3887$^\circ$ vs.\ 0.4640$^\circ$) and a reduced final standard deviation (0.1663$^\circ$ vs.\ 0.1915$^\circ$). While the maximum final error remains comparable (1.1217$^\circ$ vs.\ 1.1207$^\circ$), these results highlight the advantage of the multiplicative formulation in capturing the inherently rotational nature of sensor pointing errors. 

Moreover, these results validate the TRIAD-enhanced MEKF as a reliable and high-precision estimator. This robust performance is critical, as this filter serves as the computational backbone for each hypothesis within the MMAE framework used to address the sensor misalignment problem.

\subsection{Multiple-Model Adaptive Estimation Results and Analysis for Misalignment Estimation.}
\label{sec:single_mis_results}
The performance of the proposed MMAE framework was evaluated through a $N_{MC} = 100$ runs Monte Carlo analysis to ensure statistical robustness. Each run was initialized with the same filter parameters, including the initial covariance matrix $\mat{P}_0$, process noise covariance $\mat{Q}$, and measurement noise covariance $\mat{R}$, as summarized in Table~\ref{tab:covariance_matrices}.

To provide a comprehensive measure of the filter's performance, the Root Mean Square Error (RMSE) \cite{servadio2025dynamicalupdatemapsparticle} for attitude ($\Xi_{q,k}$), angular velocity ($\Xi_{\omega,k}$), gyro bias ($\Xi_{b,k}$), and misalignment ($\Xi_{\mu,k}$) are calculated as
\begin{align}
    \Xi_{q,k} &= \sqrt{\frac{1}{N_{MC}} \sum_{i=1}^{N_{MC}} \left( \mathbf{q}_{T,k}^{(i)} - \hat{\mathbf{q}}_{k}^{(i)+} \right)^T \left( \mathbf{q}_{T,k}^{(i)} - \hat{\mathbf{q}}_{k}^{(i)+} \right)} \label{eq:rmse_q} \\
    \Xi_{\omega,k} &= \sqrt{\frac{1}{N_{MC}} \sum_{i=1}^{N_{MC}} \left( \boldsymbol{\omega}_{T,k}^{(i)} - \hat{\boldsymbol{\omega}}_{k}^{(i)+} \right)^T \left( \boldsymbol{\omega}_{T,k}^{(i)} - \hat{\boldsymbol{\omega}}_{k}^{(i)+} \right)} \label{eq:rmse_w} \\
    \Xi_{b,k} &= \sqrt{\frac{1}{N_{MC}} \sum_{i=1}^{N_{MC}} \left( \mathbf{b}_{T,k}^{(i)} - \hat{\mathbf{b}}_{k}^{(i)+} \right)^T \left( \mathbf{b}_{T,k}^{(i)} - \hat{\mathbf{b}}_{k}^{(i)+} \right)} \label{eq:rmse_b} \\
    \Xi_{\mu,k} &= \sqrt{\frac{1}{N_{MC}} \sum_{i=1}^{N_{MC}} \left( \boldsymbol{\mu}_{T,k}^{(i)} - \hat{\boldsymbol{\mu}}_{k}^{(i)+} \right)^T \left( \boldsymbol{\mu}_{T,k}^{(i)} - \hat{\boldsymbol{\mu}}_{k}^{(i)+} \right)} \label{eq:rmse_mu}
\end{align}

\begin{figure}[H]
    \centering
    \includegraphics[width=0.95\textwidth]{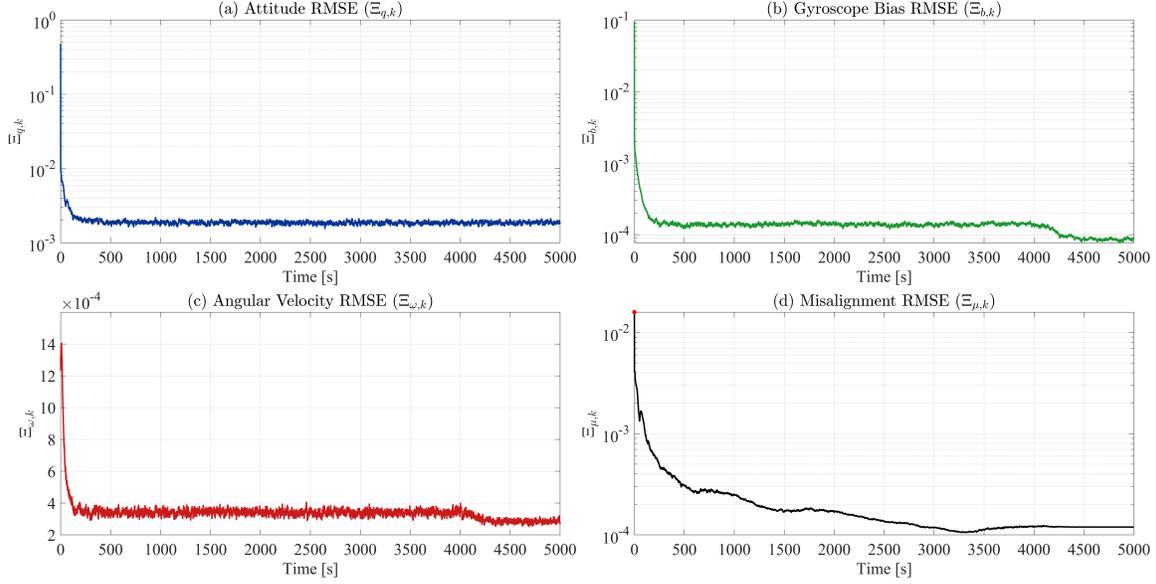}%
    \caption{Monte Carlo Root Mean Square Error (RMSE) histories for (a) attitude, (b) gyro bias, (c) angular velocity, 
    and (d) star–tracker misalignment, averaged over $N_{\mathrm{MC}} = 100$ runs.}
    \label{fig:refinement_psi_dual_trigger}
\end{figure}

The general convergence of the filter is quantified in Figure \ref{fig:refinement_psi_dual_trigger}. Each subplot demonstrates that the estimation error for its respective state rapidly decreases from the initial uncertainty and settles at a stable, low-error value.  The attitude RMSE~(a) settles to approximately $2 \times 10^{-3}$, while the star tracker misalignment RMSE~(d) converges to $1.5 \times 10^{-4}$, validating the framework's ability to identify the unknown sensor error. The gyroscope bias~(b) and angular velocity~(c) errors also showed a robust convergence towards low values.

Table~\ref{tab:mmae_results} provides a final snapshot of the component-wise mean errors at the end of the simulation horizon ($t=5000$~s), further confirming the high-fidelity performance of the filter. The results confirm arc-second accuracy, with final attitude and misalignment errors on the order of hundredths of a degree per axis. Furthermore, the final angular velocity and gyroscope bias errors are approximately $10^{-5}$~rad/s and $10^{-6}$~rad/s, respectively, demonstrating the micro-radian-level precision required for demanding deep-space applications.

\begin{table}[h]
\small
\captionsetup{width=\textwidth}
\centering
\caption{Multiple-Model Adaptive Estimation: Final Mean Errors at $t=5000.0$~s over 100 Monte Carlo runs.}
\begin{tabular}{@{} p{3.2cm} c r @{\hskip 0.5cm} p{3.2cm} c r @{}}
\toprule
\textbf{Error Metric} & \textbf{Axis} & \textbf{Mean Error} & \textbf{Error Metric} & \textbf{Axis} & \textbf{Mean Error} \\
\midrule
\label{tab:mmae_results}
\multirow{3}{=}{Attitude Error (deg)} 
 & X & 0.007101 & \multirow{3}{=}{Angular Velocity Error (rad/s)} 
 & X & $1.237 \times 10^{-5}$ \\
 & Y & 0.006030 &  & Y & $-2.93 \times 10^{-6}$ \\
 & Z & -0.021401 &  & Z & $-1.237 \times 10^{-5}$ \\
\midrule
\multirow{3}{=}{Gyroscope Bias Error (rad/s)} 
 & X & $3.6 \times 10^{-6}$ & \multirow{3}{=}{Misalignment Error (deg)} 
 & X & -0.001213 \\
 & Y & $6.98 \times 10^{-6}$ &  & Y & 0.000469 \\
 & Z & $-2.4 \times 10^{-7}$ &  & Z & 0.001374 \\
\bottomrule
\end{tabular}
\end{table}

\begin{figure}[H]
    \centering

    \begin{minipage}[t]{0.49\textwidth}
        \centering
        \textbf{(a)} Attitude component estimation errors \\
        \includegraphics[width=\linewidth]{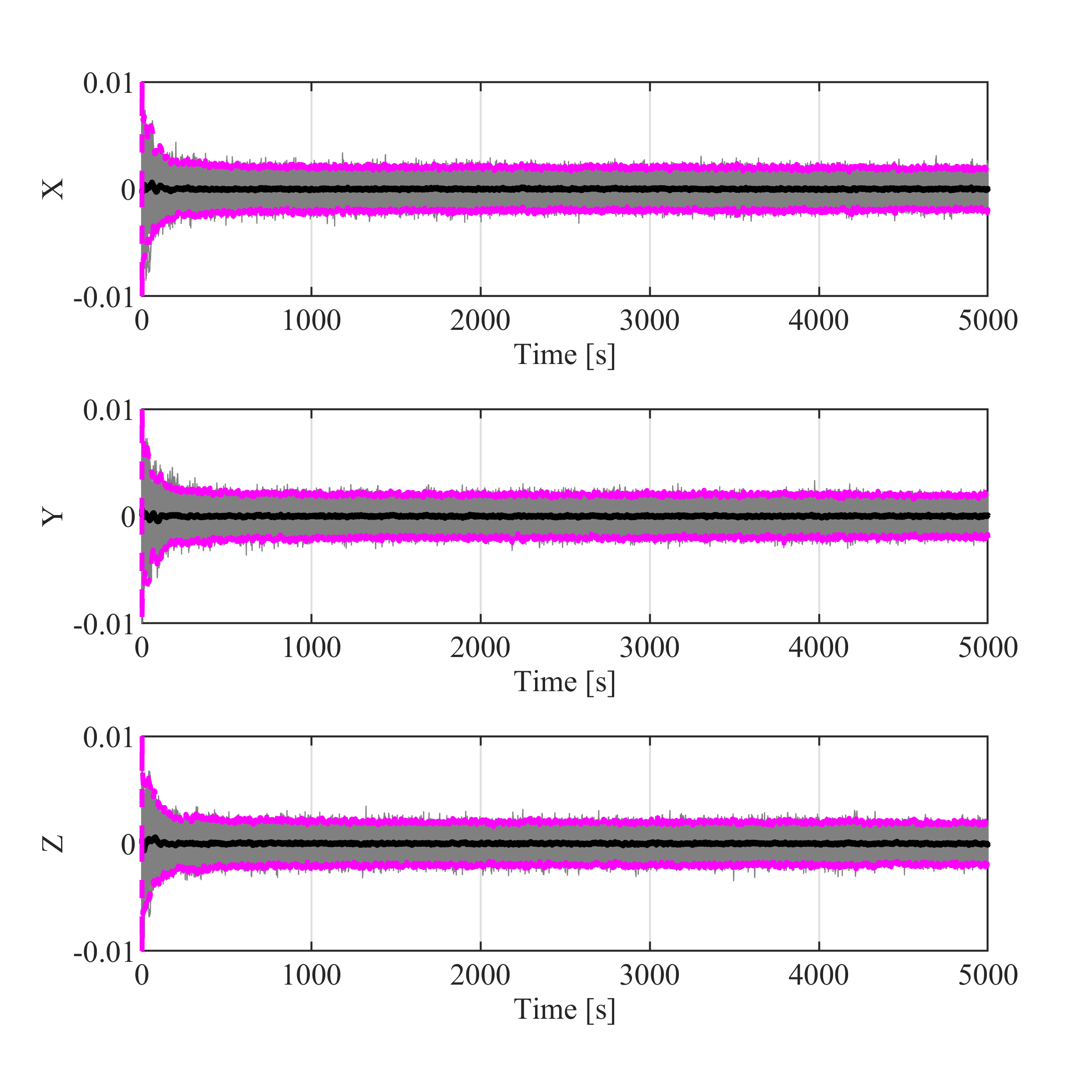}
    \end{minipage}
    \hfill
    \begin{minipage}[t]{0.482\textwidth}
        \centering
        \textbf{(b)} Gyroscope bias estimation errors \\
        \includegraphics[width=\linewidth]{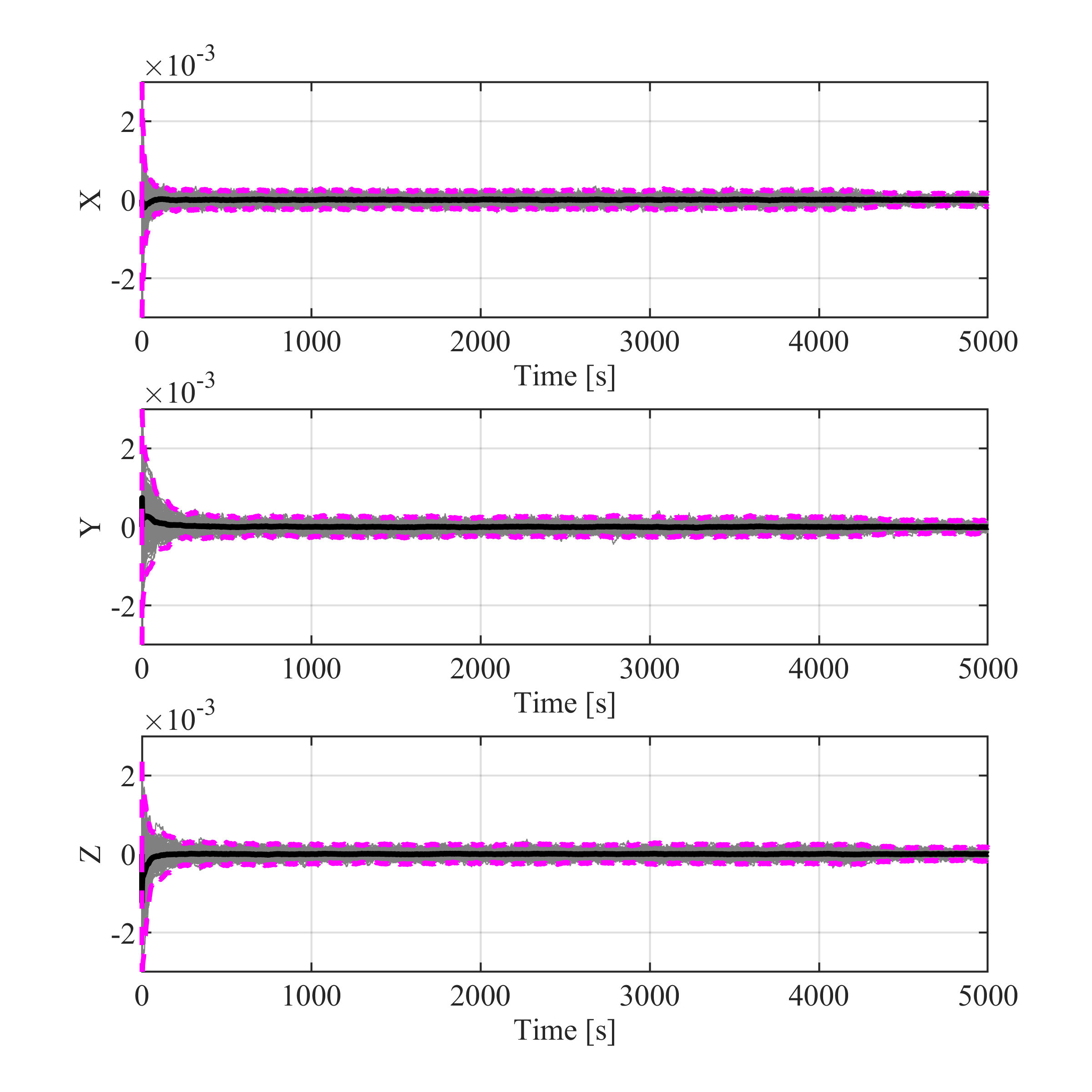}
    \end{minipage}

    \vspace{4mm}

    \begin{minipage}[t]{0.49\textwidth}
        \centering
        \textbf{(c)} Angular velocity estimation errors \\
        \includegraphics[width=\linewidth]{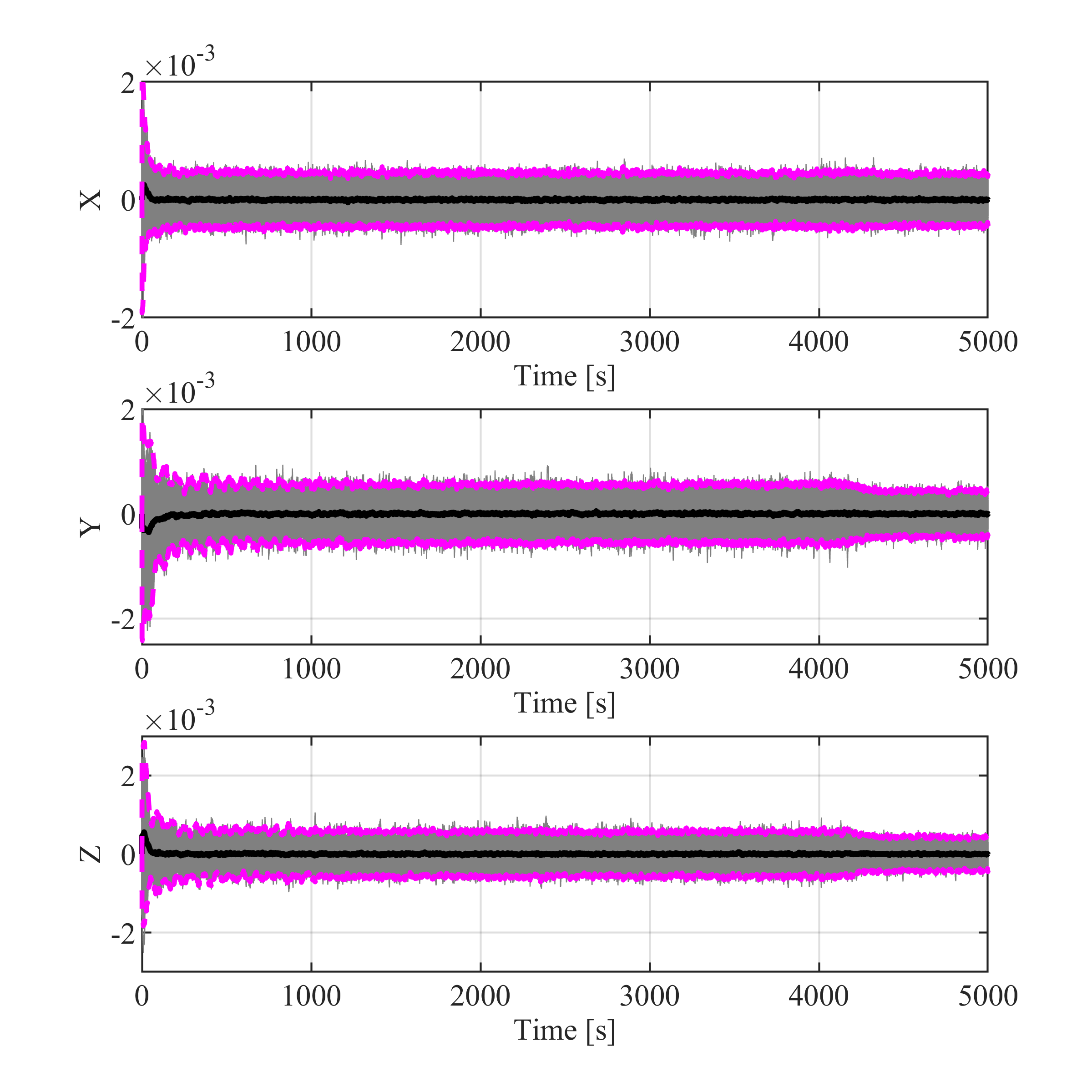}
    \end{minipage}
    \hfill
    \begin{minipage}[t]{0.488\textwidth}
        \centering
        \textbf{(d)} Star tracker misalignment estimation errors \\
        \includegraphics[width=\linewidth]{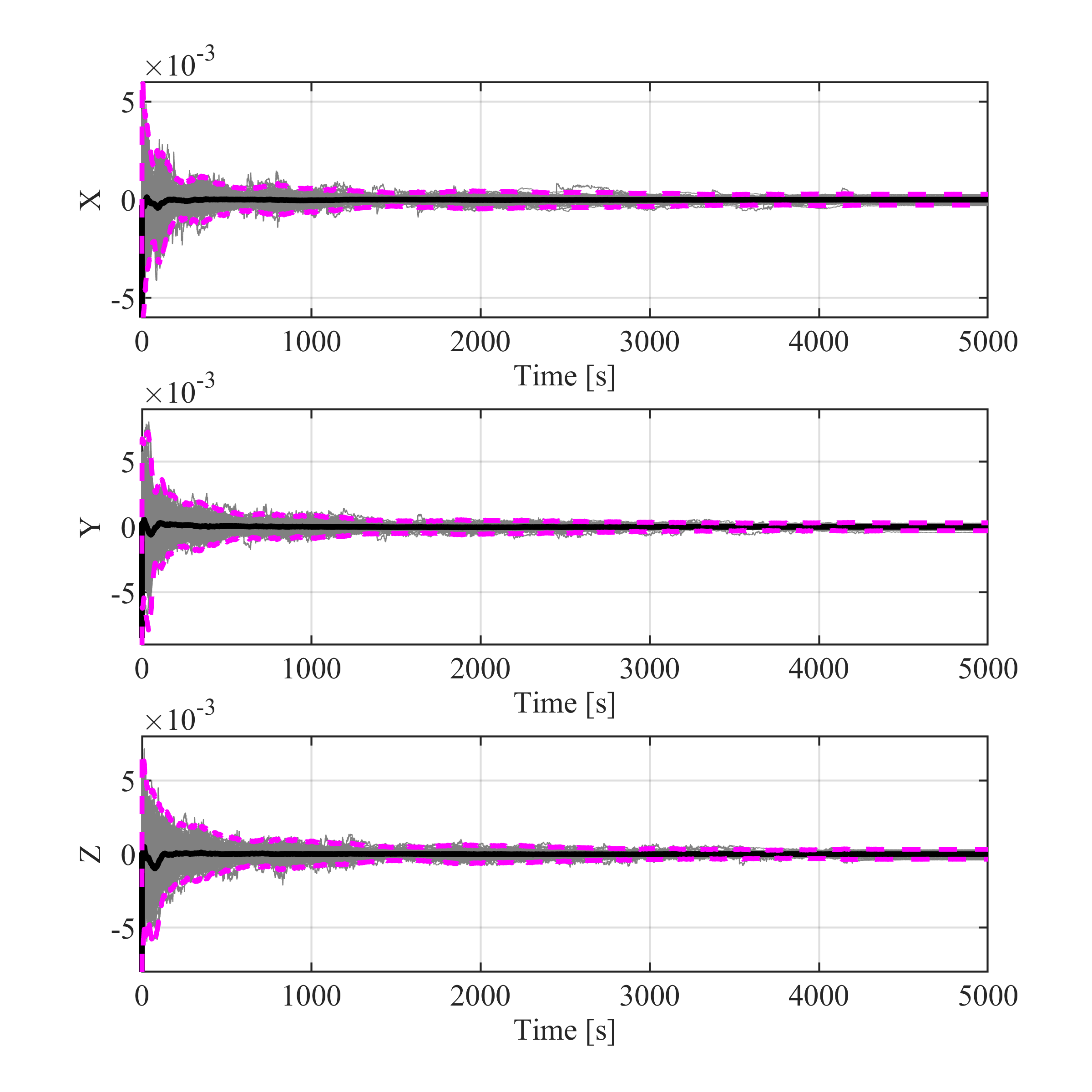}
    \end{minipage}

    \caption{
        Estimation errors over time from Monte Carlo simulations. 
        (a) Attitude estimation errors using quaternion components ($q_x$, $q_y$, $q_z$). 
        (b) Gyroscope bias estimation errors ($b_x$, $b_y$, $b_z$). 
        (c) Angular velocity estimation errors ($\omega_x$, $\omega_y$, $\omega_z$). 
        (d) Star-Tracker misalignment estimation errors ($\mu_{R,x}$, $\mu_{R,y}$, $\mu_{R,z}$). 
        All plots include $\pm3\sigma$ bounds across 100 Monte Carlo trials.
    }
    \label{fig:4panel_errors}
\end{figure}

For further validation, a filter consistency analysis was also performed. The filter's ability to jointly estimate the spacecraft attitude and sensor misalignment is demonstrated and reported in Figure~\ref{fig:4panel_errors}, which presents the time evolution of the estimation errors for all state components. The plots illustrate that the errors for attitude, gyroscope bias, angular velocity, and star tracker misalignment all converge rapidly and remain within the statistically expected $\pm3\sigma$ covariance bounds. This confirms that the filter is well-tuned and provides a consistent and reliable state estimate.

\subsection{Comparative Analysis of Adaptive Refinement Strategies}

\begin{figure}[H]
    \centering
    \includegraphics[width=1\textwidth]{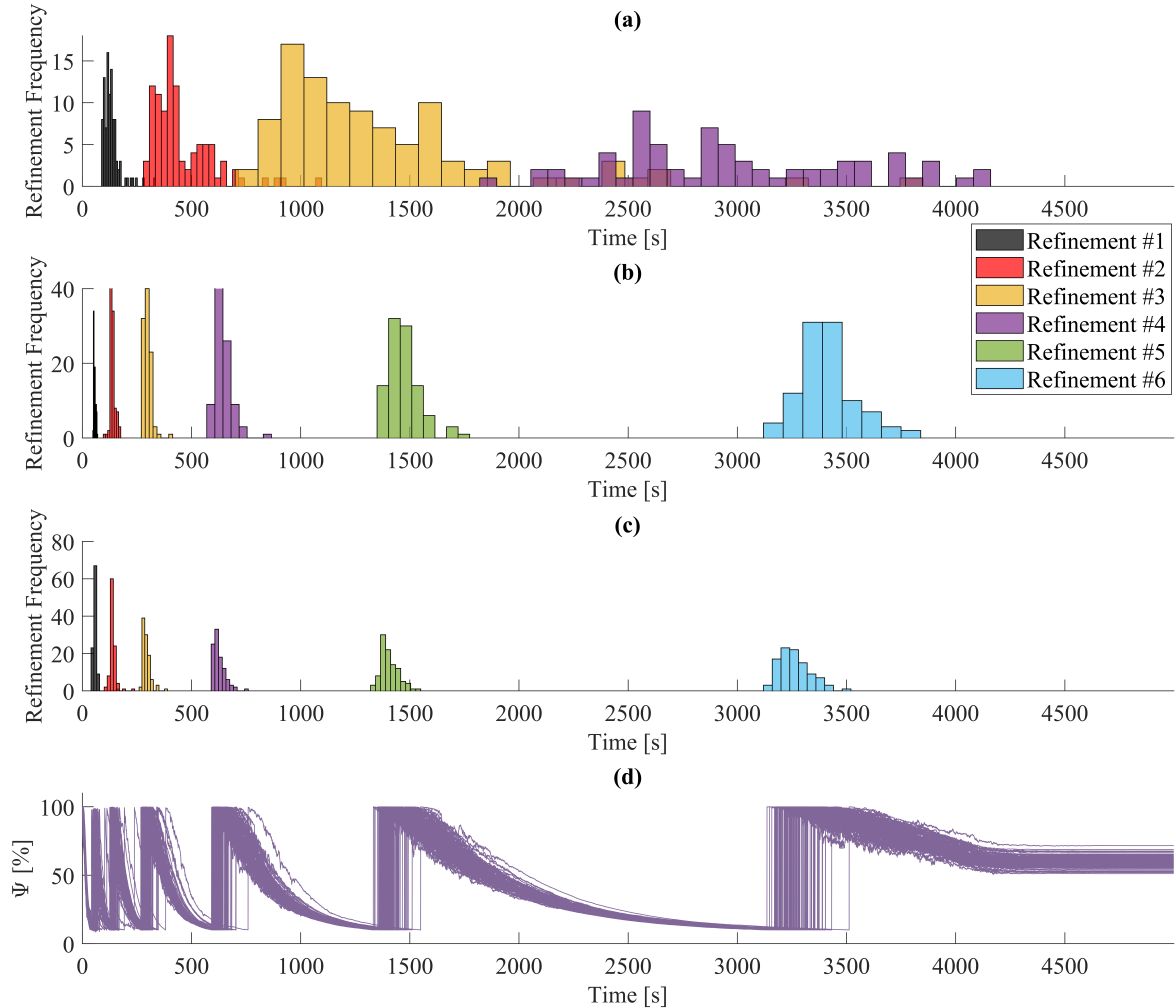}
   \caption{%
        Comparison of adaptive refinement strategies and resulting hypothesis diversity.
        \textbf{(a):} Classical Trigger (MAP Center). 
        \textbf{(b):} $\Psi$-Based Trigger (MAP Center). 
        \textbf{(c):} $\Psi$-Based Trigger (mean Center). 
        \textbf{(d):} Evolution of the Hypothesis Diversity Metric, $\Psi$, over 100 Monte Carlo runs for the $\Psi$-Based Trigger (mean Center) case.
    }
    \label{fig:refinement_comparison} 
\end{figure}

The behavior of the adaptive refinement strategies is illustrated in Figure~\ref{fig:refinement_comparison}. The upper three subplots present a histogram of the refinement event times across 100 Monte Carlo runs, while the lower subplot, Figure~\ref{fig:refinement_comparison}(d), displays the time evolution of the Hypothesis Diversity Metric $\Psi$, corresponding to the strategy shown in Figure~\ref{fig:refinement_comparison}(c), the third approach. The $\Psi$-based methodology continues to refine until the grid of hypotheses reaches below measurement noise levels, at which point it becomes impossible for the MMAE to prefer one model over the others. The refinement mechanics cease when the MMAE reaches its steady state, indicated by the constant values, with approximately 50\% to 80\% of models being effectively active, thereby providing no preferred refinement direction. 

The classical weight-based trigger is the least effective strategy due to premature convergence, stalling at a final RMSE of \( 1.999 \times 10^{-4} \) after only 3.69 average refinements. This failure occurs because the trigger requires a single model to exceed a 50\% weight threshold. Late in the simulation, as the overall model diversity $\Psi$ decreases due to a cluster of good hypotheses, no single model becomes dominant enough to trigger the refinement threshold. Therefore, the trigger mechanism traps the filter in a sub-optimal state, preventing the final refinements needed to eliminate remaining errors.

\begin{figure}[H]
    \centering
    \includegraphics[width=0.85\textwidth]{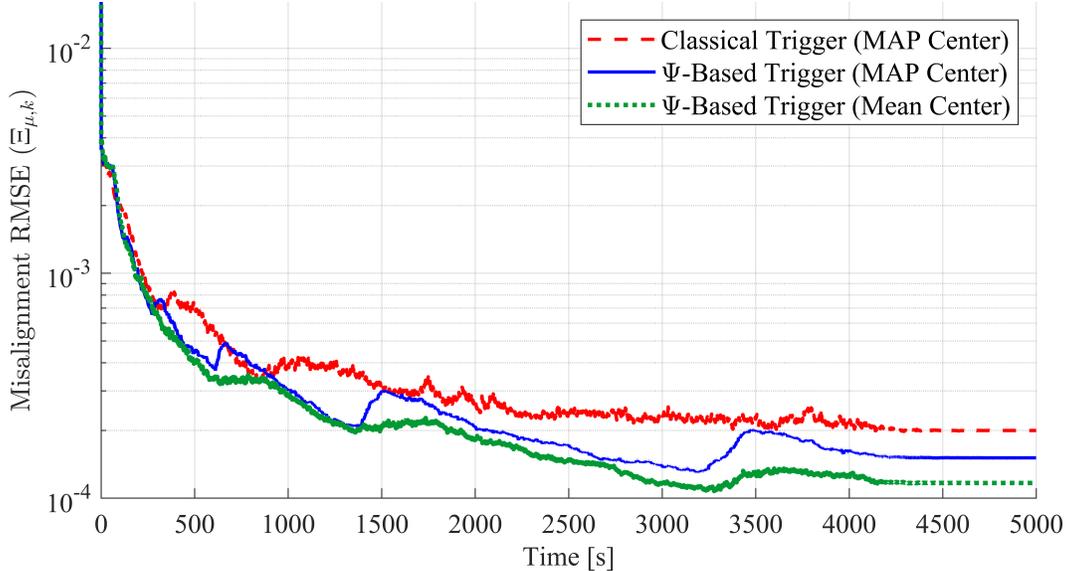}
    \caption{Comparison of misalignment RMSE ($\Xi_{\mu,k}$) between the proposed $\Psi(t)$-based trigger and the classical weight-based trigger. The proposed method achieves a 41.57\% reduction in final error by initiating additional refinements where the classical method stalls.}
    \label{fig:rmse_comparison}
\end{figure}

The advantage of the $\Psi$-based triggers is their ability to overcome the premature convergence of the classical method, as shown in Fig.~\ref{fig:rmse_comparison}. These triggers initiate further refinements late in the simulation (near $t=1500$~s and $t=3300$~s), driving further reductions in misalignment error. Although both $\Psi$-based strategies use, on average, the same number of refinements, the weighted mean centering approach is demonstrably superior. In the MAP-centered strategy, the refinement grid is re-centered on the single hypothesis with the highest weight at that instant. When the posterior is still multi-modal, small fluctuations in the dominant model can cause the grid center to jump between neighboring hypotheses, producing the bumps visible in the MAP curve of Fig.~\ref{fig:rmse_comparison}. In contrast, the mean-centered trigger uses the full posterior, re-centering the grid on the weighted average of all hypotheses. This yields a smoother, more stable motion of the grid center, keeps refinements aligned with the bulk of the probability mass, and eliminates transient bumps, resulting in the lowest final misalignment RMSE of $1.168\times 10^{-4}$~rad ($\approx 24.08$~arcsec).

The analysis confirms that the proposed $\Psi$-based trigger with weighted mean centering provides the most accurate and robust performance. By preventing premature convergence and leveraging the full posterior distribution for grid refinement, this strategy reduces the final misalignment RMSE by 24.42\% compared to the $\Psi$-based MAP-centered approach and by a significant 41.57\% compared to the classical triggering approach. This demonstrates the method's robustness and confirms its selection as the optimal approach for this high-fidelity estimation problem.

\begin{table}[tbh]
\centering
\caption{Final Star-Tracker Misalignment RMSE ($\Xi_{\mu,k}$) and Average Refinement Counts for all Strategies over 100 Monte Carlo runs. }
\label{tab:avg_refinements}
\begin{tabular}{@{} l r r @{}} 
\toprule
\textbf{Strategy} & \textbf{Average Refinements} & \textbf{Final RMSE ($\Xi_{\mu,k}$)}(radians) \\
\midrule
Classical Trigger (MAP Center)              & 3.69 & $1.999 \times 10^{-4}$ \\
$\Psi$-Based Trigger (MAP Center)           & 6.00 & $1.511 \times 10^{-4}$ \\
$\Psi$-Based Trigger (Weighted Mean Center) & 6.00 & $1.168 \times 10^{-4}$ \\
\bottomrule
\end{tabular}
\end{table}

\section{Dual Misalignment Estimation Problem}

Many deep-space CubeSats feature two star-trackers to increase sky coverage, provide redundancy, and obtain a larger field of view of the night sky, at the cost of introducing two independent sensor misalignments relative to the body frame. In this section, we extend the single-misalignment estimation problem to the dual-misalignment configuration, allowing both star trackers to be calibrated simultaneously without increasing the continuous state dimension of the filter. This enables a more robust and accurate attitude solution for deep-space CubeSat missions.

\subsection{Problem Formulation: Double Misalignment Scenario}
\label{dualmisalignmentprob}
Relative to the single-misalignment case, the MEKF structure, continuous time error dynamics $F$, discretization $\Phi$, and process/measurement noise models are identical to the single-misalignment formulation. The MMAE hypothesis becomes six-dimensional (6D), evaluating pairs
$\{(\boldsymbol{\mu}_1^{(j)}, \boldsymbol{\mu}_2^{(j)})\}_{j=1}^N$
with a joint likelihood over all stars and gyro residuals.
 All other filter operations (weight normalization, pruning/refinement, grid centering, and quaternion fusion) remain unchanged \cite{ganganath2025startrackermisalignmentcompensation}.

\begin{figure}[H]
  \centering
  \includegraphics[width=0.7\linewidth]{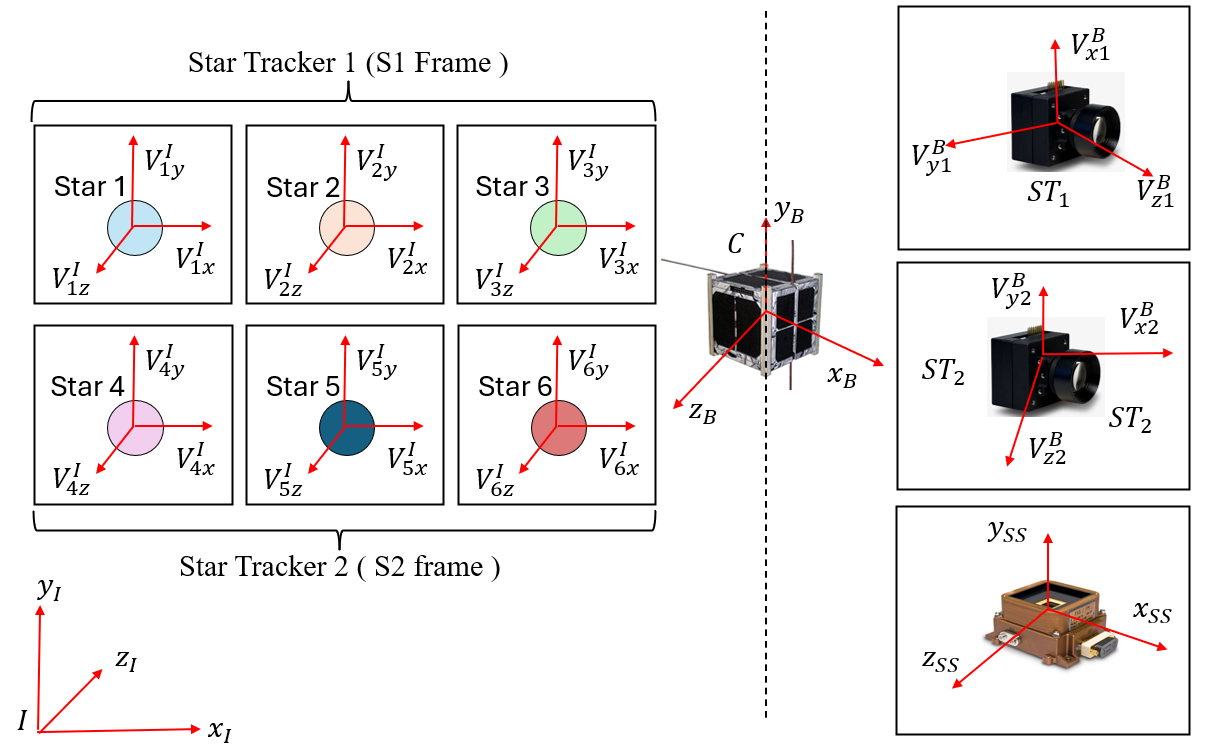}
  \caption{Coordinate frames for the CubeSat (dual misalignment case),
  including the inertial ($\mathcal{I}$), body ($\mathcal{B}$), gyroscope
  ($\mathcal{G}$), and two star-tracker ($ST_1$, $ST_2$) reference frames.
  Camera~1 observes three inertial stars
  $\{v^{I}_{1}, v^{I}_{2}, v^{I}_{3}\}$ and Camera~2 observes
  $\{v^{I}_{4}, v^{I}_{5}, v^{I}_{6}\}$. Each panel shows the sensor FOV on
  the unit sphere and the corresponding LOS directions from the CubeSat body
  frame.}
  \label{fig:frames}
\end{figure}

In this problem, we consider a rigid spacecraft equipped with two star-trackers, $\mathrm{ST}_1$ and $\mathrm{ST}_2$, each mounted on the CubeSat with its own small, fixed misalignment relative to the body frame
$\mathcal{B} = (x_B, y_B, z_B)$. The star–tracker sensor frames are denoted $\mathcal{S}_1$ and $\mathcal{S}_2$, and the corresponding body-to-sensor misalignments are parameterized by small-angle vectors $\boldsymbol{\mu}_1, \boldsymbol{\mu}_2 \in \mathbb{R}^3$. For each camera $k \in \{1,2\}$, we associate to $\boldsymbol{\mu}_k$ a unit misalignment quaternion $\delta \boldsymbol{q}(\boldsymbol{\mu}_k)$ and the corresponding direction–cosine matrix

\begin{equation}
\boldsymbol{C}_{\mu_k} \;=\; \boldsymbol{C}\big(\delta \boldsymbol{q}(\boldsymbol{\mu}_k)\big) \in \mathrm{SO}(3)
\end{equation}
where $\boldsymbol{C}(\cdot)$ denotes the standard mapping from a unit quaternion to a $3\times 3$ rotation matrix. Thus $\boldsymbol{C}_{\mu_k}$ rotates body-frame vectors into the $k$th sensor frame. The true spacecraft attitude is described by the inertial-to-body quaternion $\boldsymbol{q}(t)$, with its corresponding direction–cosine matrix defined as $\boldsymbol{C}_{\bm q}$, which rotates inertial-frame vectors into the body frame. In this simulation, it is assumed that each tracker observes three known inertial unit vectors (three stars). For $\mathrm{ST}_1$, the set $\{\boldsymbol{v}^{I}_{1}, \boldsymbol{v}^{I}_{2}, \boldsymbol{v}^{I}_{3}\}$ is selected, and for $\mathrm{ST}_2$, the set $\{\boldsymbol{v}^{I}_{4}, \boldsymbol{v}^{I}_{5}, \boldsymbol{v}^{I}_{6}\}$ is selected, all expressed in the inertial frame $\mathcal{I} = (x_I, y_I, z_I)$.
 The ideal line-of-sight (LOS) directions in each sensor frame are denoted as
\begin{equation}
  \boldsymbol{v}^{(k)}_{S,i,\text{true}} \;=\; \boldsymbol{C}_{\mu_k}\,\boldsymbol{C}_{\bm q}\,\boldsymbol{v}^{I}_{i},
  \qquad
  \begin{cases}
    k = 1,\; i \in \{1,2,3\},\\
    k = 2,\; i \in \{4,5,6\},
  \end{cases}
\end{equation}
where the ``S" subscript indicates that the vector is expressed with respect to the $k$-th star tracker frame, being rotated by the CubeSat quaternion DCM and the misalignment DCM, in series. The measured, i.e., affected by stochastic noise, LOS vectors are
\begin{equation}
  \boldsymbol{y}^{(k)}_{i} \;=\; \boldsymbol{v}^{(k)}_{S,i,\text{true}} + \boldsymbol{\eta}^{(k)}_{i},
  \qquad
  \boldsymbol{\eta}^{(k)}_{i} \sim \mathcal{N}(\mathbf{0}, \boldsymbol{R}_{\boldsymbol{\eta}\boldsymbol{\eta}}),
\end{equation}
with independent zero-mean Gaussian noise $\boldsymbol{\eta}^{(k)}_{i}$ on each vector measurement. The rate gyroscope, whose axes define frame $\mathcal{G}$ (assumed to be nearly aligned with the body frame), reports the body angular velocity corrupted by an additive bias and measurement noise,
\begin{equation}
  \boldsymbol{y}_{\boldsymbol{\omega}} \;=\; \boldsymbol{\omega} + \boldsymbol{b} + \boldsymbol{\eta}_{\boldsymbol{\omega}}
  \qquad
  \boldsymbol{\nu}_{\boldsymbol{\omega}} \sim \mathcal{N}(\mathbf{0}, \boldsymbol{R}_{{\boldsymbol{\omega}}{\boldsymbol{\omega}}}),
  \label{eq:gyro_meas_model}
\end{equation}
where $\boldsymbol{\omega}$ is the true body angular velocity, $\boldsymbol{b}$ is a slowly varying gyro bias, and $\boldsymbol{\eta}_{\boldsymbol{\omega}}$ is zero–mean white noise with covariance $\boldsymbol{R}_{\boldsymbol{\omega}\boldsymbol{\omega}}$. The goal is to jointly estimate the time-varying attitude $\boldsymbol{q}(t)$, angular velocity $\boldsymbol{\omega}(t)$, and gyro bias $\boldsymbol{b}(t)$, together with the two constant misalignment vectors $\boldsymbol{\mu}_1$ and $\boldsymbol{\mu}_2$. To this end, we adopt the previously derived MEKF for the error state $[\delta\bm\omega^\top,\;\delta\bm b^\top,\;\delta \bm\theta^\top]^\top$ and while the MMAE evaluates a six-dimensional grid of hypotheses $\{(\boldsymbol{\mu}_1^{(j)},\,\boldsymbol{\mu}_2^{(j)})\}_{j=1}^{N}$. The corresponding dual camera geometry and star tracker selection are illustrated in Fig.~\ref{fig:frames}.

For a small MRP vector $\delta\bm{\theta}$, the corresponding DCM admits the first–order approximation 
\begin{equation}\label{eq:small_rot}
  \bm C_{\delta\bm\theta} = \bm C\big(\delta \bm q(\delta\bm{\theta})\big)
  \;=\;
  \I_3 \;+\; 4\,\crs{\delta\bm{\theta}} \;+\; \mathcal{O}(\|\delta\bm{\theta}\|^2),
\end{equation}
since for MRPs the rotation angle satisfies
$\alpha = 4\arctan(\|\delta\bm{\theta} \|) \approx 4\|\delta\bm{\theta}\|$. Under the left-multiplicative small attitude error parameterized by the MRP vector $\delta \bm \theta$, such that 
$\bm q_{\text{true}}=\delta \bm q(\delta \bm \theta)\otimes\bm{\hat q}$, the corresponding measurement vectors, in terms of LOS directions, are evaluated as 
\begin{equation}
  \hat{\bm v}^{(k)}_{S,i} =\boldsymbol{C}_{\bm \mu_k}\,\bm C_{\delta\bm\theta}\boldsymbol{C}_{{\bm q}}\,\boldsymbol{v}^{I}_{i},
  \qquad
  \begin{cases}
    k = 1,\; i \in \{1,2,3\},\\
    k = 2,\; i \in \{4,5,6\},
  \end{cases}
\end{equation}
which highlights the three rotations in series: starting from the estimated quaternion, followed by the small MRP for each MEKF, and ending with the misalignment rotation for each of the MMAE models.

Using the standard skew-symmetric and rotation identities \cite{MarkleyCrassidis2014,MurrayLiSastry1994,LynchPark2017ModernRobotics}, we can linearize the star-vector measurement about the current attitude estimate to obtain the Jacobians required by the MEKF in the covariance evaluations. This makes the dependence on $\delta \bm \theta$ explicit and yields closed-form Jacobians. Considering the standard residual for the $i$th vector and $k$th misalignment 
\begin{align}\label{eq:rk_linear}
\bm r_i^{(k)}&=\bm y_i^{(k)}-\hat{\bm v}^{(k)}_{S,i} \nonumber\\
&= \Big(\boldsymbol{C}_{\bm \mu_k}\big(\I_3+4[\delta\bm\theta]_\times\big)\boldsymbol{C}_{{\bm q}}\,\boldsymbol{v}^{I}_{i}
       - \boldsymbol{C}_{\bm \mu_k}\boldsymbol{C}_{{\bm q}}\,\boldsymbol{v}^{I}_{i}\Big) + \boldsymbol{\eta}^{(k)}_{i} \nonumber\\
&= 4\,\boldsymbol{C}_{\bm \mu_k}\,[\delta\bm\theta]_\times\,\boldsymbol{C}_{{\bm q}}\,\boldsymbol{v}^{I}_{i} + \boldsymbol{\eta}^{(k)}_{i} \nonumber \\
&= -\,4\,\boldsymbol{C}_{\bm \mu_k}\,[\boldsymbol{C}_{{\bm q}}\,\boldsymbol{v}^{I}_{i}]_\times\,\delta\bm\theta + \boldsymbol{\eta}^{(k)}_{i} \nonumber \\
&= -\,4\,[\hat{\bm v}^{(k)}_{S,i}]_\times\,\boldsymbol{C}_{\bm \mu_k}\,\delta\bm\theta + \boldsymbol{\eta}^{(k)}_{i}
\end{align}
where we used the identities $[\bm a]_\times \bm b = -[\bm b]_\times \bm a$ and $C[\bm a]_\times = -[C\bm a]_\times C$ \cite{MarkleyCrassidis2014,MurrayLiSastry1994,LynchPark2017ModernRobotics}, makes the dependence on $\delta\bm\theta$ explicit and yields closed-form Jacobians. Differentiating with respect to the MEKF error state gives the following Jacobian for each star measurement: 
\begin{equation}\label{eq:drds}
\frac{\partial \bm r_i^{(k)}}{\partial \delta\bm\theta} \;=\; -\,4\,[\hat{\bm v}^{(k)}_{S,i}]_\times\,\boldsymbol{C}_{\bm \mu_k},
\qquad
\frac{\partial \bm r_i^{(k)}}{\partial \delta\bm\omega}=\Zero_{3\times 3},\quad
\frac{\partial \bm r_i^{(k)}}{\partial \delta\bm b}=\Zero_{3\times 3}.
\end{equation}

Each camera contributes three linearly independent attitude directions, stacking the per-star Jacobians from \eqref{eq:drds} yields nine measurement rows per camera. For the first star tracker with inertial stars $\bm v_{I,1},\bm v_{I,2},\bm v_{I,3}$, the measurement Jacobian is
\begin{equation}\label{eq:H_cam1}
H_{\text{cam1}} \;=\;
\begin{bmatrix}
\Zero & \Zero & -4\,\crs{\hat{\bm v}^{(1)}_{S,1}}\,\boldsymbol{C}_{\bm \mu_1}\\
\Zero & \Zero & -4\,\crs{\hat{\bm v}^{(1)}_{S,2}}\,\boldsymbol{C}_{\bm \mu_1}\\
\Zero & \Zero & -4\,\crs{\hat{\bm v}^{(1)}_{S,3}}\,\boldsymbol{C}_{\bm \mu_1}
\end{bmatrix}.
\end{equation}
Similarly, for the second star tracker observing $\bm v_{I,4},\bm v_{I,5},\bm v_{I,6}$,
\begin{equation}\label{eq:H_cam2}
H_{\text{cam2}} \;=\;
\begin{bmatrix}
\Zero & \Zero & -4\,\crs{\hat{\bm v}^{(2)}_{S,4}}\,\boldsymbol{C}_{\bm \mu_2}\\
\Zero & \Zero & -4\,\crs{\hat{\bm v}^{(2)}_{S,5}}\,\boldsymbol{C}_{\bm \mu_2}\\
\Zero & \Zero & -4\,\crs{\hat{\bm v}^{(2)}_{S,6}}\,\boldsymbol{C}_{\bm \mu_2}
\end{bmatrix}.
\end{equation}
These $3\times 9$ per-star blocks have the standard structure
$\big[\Zero\;\;\Zero\;\;-4\,\crs{\cdot}\,\boldsymbol{C}_{\bm \mu_k}\big]$, reflecting that the star LOS rows are sensitive only to the attitude error $\delta\bm\theta$ in the left-multiplicative MEKF.

The gyroscope predicted measurement yields 
\begin{equation}\label{eq:gyro_model}
\hat{\bm y}_\omega=\hat{\bm\omega}+\hat{\bm b},
\end{equation}
with corresponding  residual  
\begin{equation}\label{eq:gyro_resid}
\bm r_\omega \;=\; \bm y_\omega - \hat{\bm y}_\omega
\;=\; \delta\bm\omega + \delta\bm b + \boldsymbol{\eta}_{\boldsymbol{\omega}}
\end{equation}
The corresponding Jacobian is
\begin{equation}\label{eq:gyro_jac}
\frac{\partial \bm r_\omega}{\partial \delta\bm\omega}=\I_3,
\qquad
\frac{\partial \bm r_\omega}{\partial \delta\bm b}=\I_3,
\qquad
\frac{\partial \bm r_\omega}{\partial \delta\bm\theta}=\Zero.
\end{equation}
Thus, the gyro contributes three rows that directly constrain
bias and angular velocity and are decoupled from the attitude error
$\delta\bm\theta$. Stacking the two cameras and the gyro yields the $21\times 9$ Jacobian used in
each MEKF measurement update
\begin{equation}\label{eq:H_full}
H \;=\;
\begin{bmatrix}
H_{\mathrm{cam1}}\\[2pt]
H_{\mathrm{cam2}}\\[2pt]
\begin{matrix} \I_3 & \I_3 & \Zero_{3\times 3}\end{matrix}
\end{bmatrix}.
\end{equation}

\begin{figure}[H]
  \centering
  \includegraphics[width=0.99\linewidth]{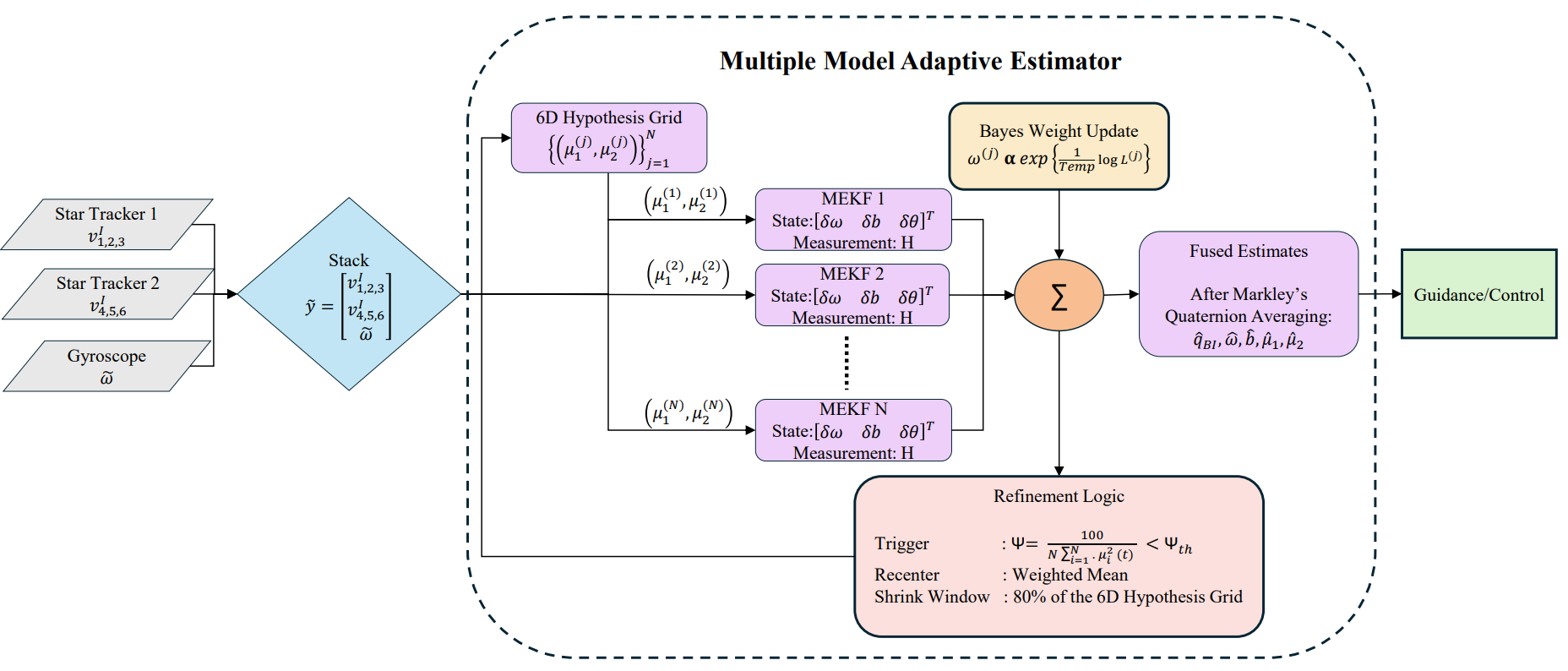}
  \caption{Two-misalignment multiple stars MEKF–MMAE block diagram.}
  \label{fig:mmae_double}
\end{figure}

The proposed theory generalizes to any number of stars identified by the star tracker without loss of generality. In the proposed numerical application, we select the star triplets for each camera so that the measurements are informative in all directions of the attitude error. The overall estimator is summarized in Fig.~\ref{fig:mmae_double}, highlighting how the 6D hypothesis grid over $(\bm \mu_1,\bm\mu_2)$ conditions a bank of MEKFs. Model likelihoods update Bayes weights, which trigger diversity-based refinement and feed Markley’s quaternion averaging to produce updated estimates via measurement fusion. That is, the likelihood evaluation and quaternion averaging is analoguos to the previous case, but adapted to the new measurement model with many star detection and multiple star trackers.

\subsection{Numerical Results}
\label{dualmisalignmentresults}
In this dual misalignment estimation problem, a $N_{MC}=100$ Monte Carlo analysis is performed to evaluate the consistency and accuracy level of the MEKF-MMAE filter. Each simulation is run for $250{,}000~\text{s}$ with a $0.5~\text{s}$ sampling period, with the application of a braking torque at $230{,}000~\text{s}$. 

To provide a challenging yet realistic and well–conditioned simulation, we select two complementary three–star constellations that yield a rich measurement geometry for the stacked Jacobian $H$. Camera~1 uses a wide southern–sky triangle (Sirius–Adhara–Canopus), while Camera~2 uses a tighter northern–sky triangle (Deneb–Sadr–Albireo). Thus, each start tracker is embedded with three stars as their known star mapping. The J2000 right ascensions, declinations, bore–to–star angles, and inertial unit vectors $v_I$ for all six stars are reported in Table~\ref{tab:st_geometry_details_with_vectors}. 

\begin{table}[H]
\centering
\caption{Per-star coordinates and angles with J2000 inertial unit vectors. “Bore$\to$star” is the angular separation from the boresight.}
\label{tab:st_geometry_details_with_vectors}
\begin{tabular}{@{} l l r r r r r r @{}}
\toprule
\textbf{Camera} & \textbf{Star} & \textbf{RA [deg]} & \textbf{Dec [deg]} & 
\textbf{Bore$\to$star [deg]} & $[v_I]_x$ & $[v_I]_y$ & $[v_I]_z$ \\
\midrule
\multirow{3}{*}{Cam 1}
& Sirius   & 101.287136 & $-16.716113$ & 18.11 &  $-0.187455$ &  $0.939218$ & $-0.287630$ \\
& Adhara   & 104.656433 & $-28.972074$ &  7.37 &  $-0.221358$ &  $0.846388$ & $-0.484383$ \\
& Canopus  &  95.987990 & $-52.695671$ & 18.11 &  $-0.063223$ &  $0.602742$ & $-0.795428$ \\
\cmidrule(l){2-8}
\multicolumn{2}{@{}l}{Pairwise separations (deg)} & \multicolumn{6}{l@{}}{12.64,\;36.22,\;24.57} \\
\midrule
\multirow{3}{*}{Cam 2}
& Deneb    & 310.357979 &  45.280339   & 11.14 &   $0.455649$ & $-0.536182$ &  $0.710558$ \\
& Sadr     & 305.164999 &  40.256708   &  4.92 &   $0.439527$ & $-0.623878$ &  $0.646213$ \\
& Albireo  & 292.680371 &  27.959678   & 11.14 &   $0.340583$ & $-0.814974$ &  $0.468850$ \\
\cmidrule(l){2-8}
\multicolumn{2}{@{}l}{Pairwise separations (deg)} & \multicolumn{6}{l@{}}{6.30,\;22.29,\;16.03} \\
\bottomrule
\end{tabular}
\end{table}

\begin{figure}[h]
    \centering
    \includegraphics[width=0.99\textwidth]{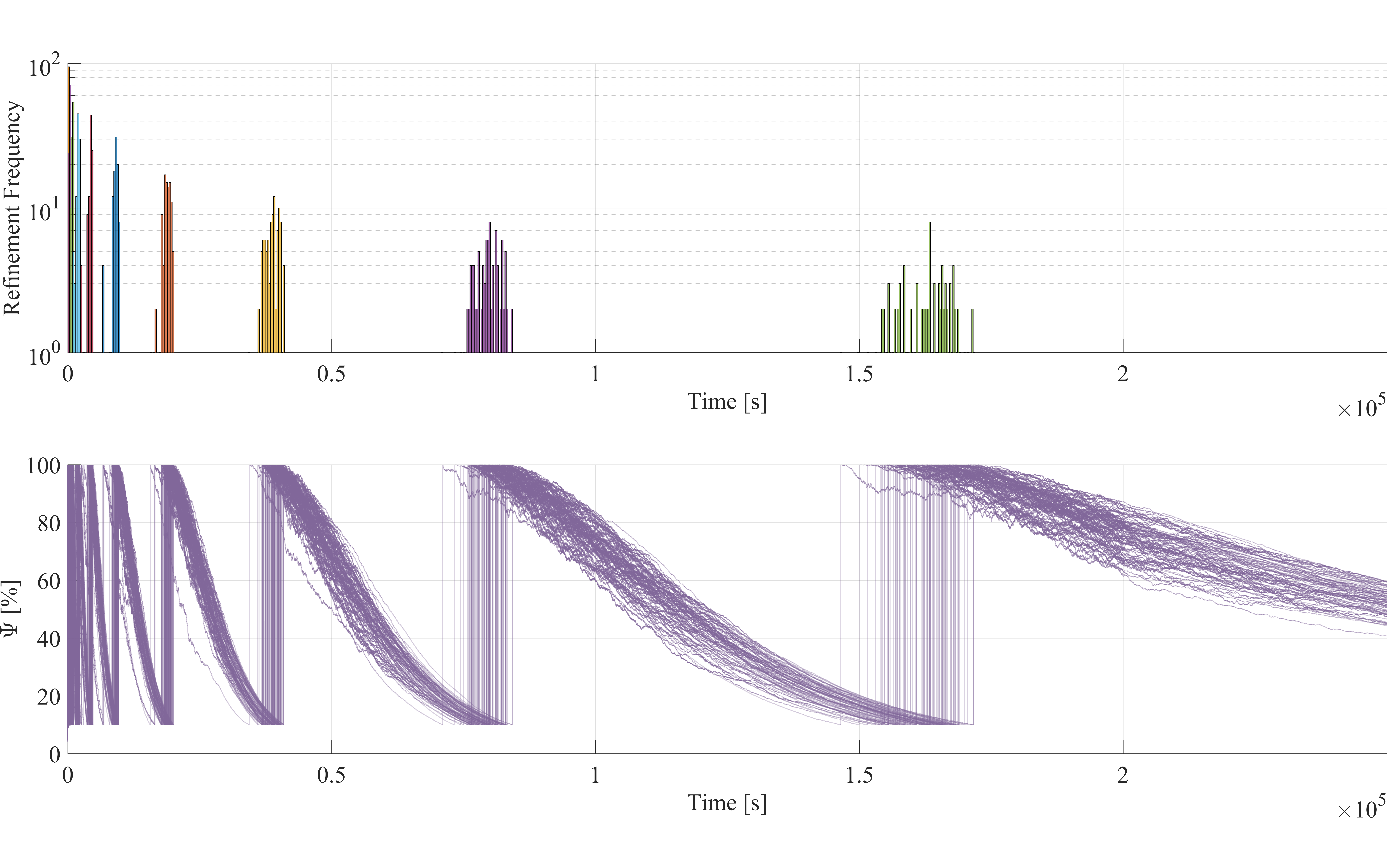}
    \caption{Refinement statistics for the dual-misalignment MMAE using the $\Psi$-based trigger with a 10\% threshold and weighted-mean grid centering.
    Top: histogram of refinement times over 100 Monte Carlo runs (each color denotes a different refinement). Bottom: evolution of the hypothesis
    diversity metric $\Psi(t)$.}
    \label{fig:dual_psi_refinement}
\end{figure}

We employ the same refinement strategy that was found to be optimal in the single-misalignment study, namely the $\Psi$-based
trigger with a 10\% threshold and weighted-mean grid centering (see Sec.~\ref{fig:rmse_comparison} and Table~\ref{tab:avg_refinements}). Figure~\ref{fig:dual_psi_refinement} summarizes the resulting refinement behavior after the Monte Carlo analysis. The top panel shows that refinements occur
in distinct bursts. As the filter bank begins from a very coarse and large 6D grid, $\Psi$ rapidly decays from near $100\%$ and repeatedly crosses the 10\% threshold, triggering early refinements clustered near the start of each run. This capability of rapidly re-centering the misalignment grid around the most likely models gives the filter robustness and overcomes the large uncertainty issue that would break the classic MEKF without the MMAE bank of models, as the initial misalignment covariance would have been too large for convergence. As the posterior over $({\bm \mu_1},{\bm \mu_2})$ sharpens, subsequent refinements occur later and are more widely spaced in time, reflecting the reduced need for aggressive grid contraction once the high-probability region has been localized. However, $\Psi$ continues decreasing over time, indicating a continuous improvement in accuracy. 

Each refinement instant produces a vertical jump in $\Psi$ back toward $100\%$ as the hypothesis grid is re-centered and reinitialized around the weighted-mean misalignment estimate. Between refinements, $\Psi$ decays smoothly as the weights concentrate on a small subset of models. As subsequent refinements are performed, $\Psi$ decays with a slower slope, as it becomes harder for the MMAE to separate the models via their likelihood due to the stochastic nature of the system. This behavior mirrors the single-misalignment case, confirming that the $\Psi$-based, mean-centered strategy scales effectively to the full 6D dual-star-tracker misalignment problem.

\begin{figure}[t]
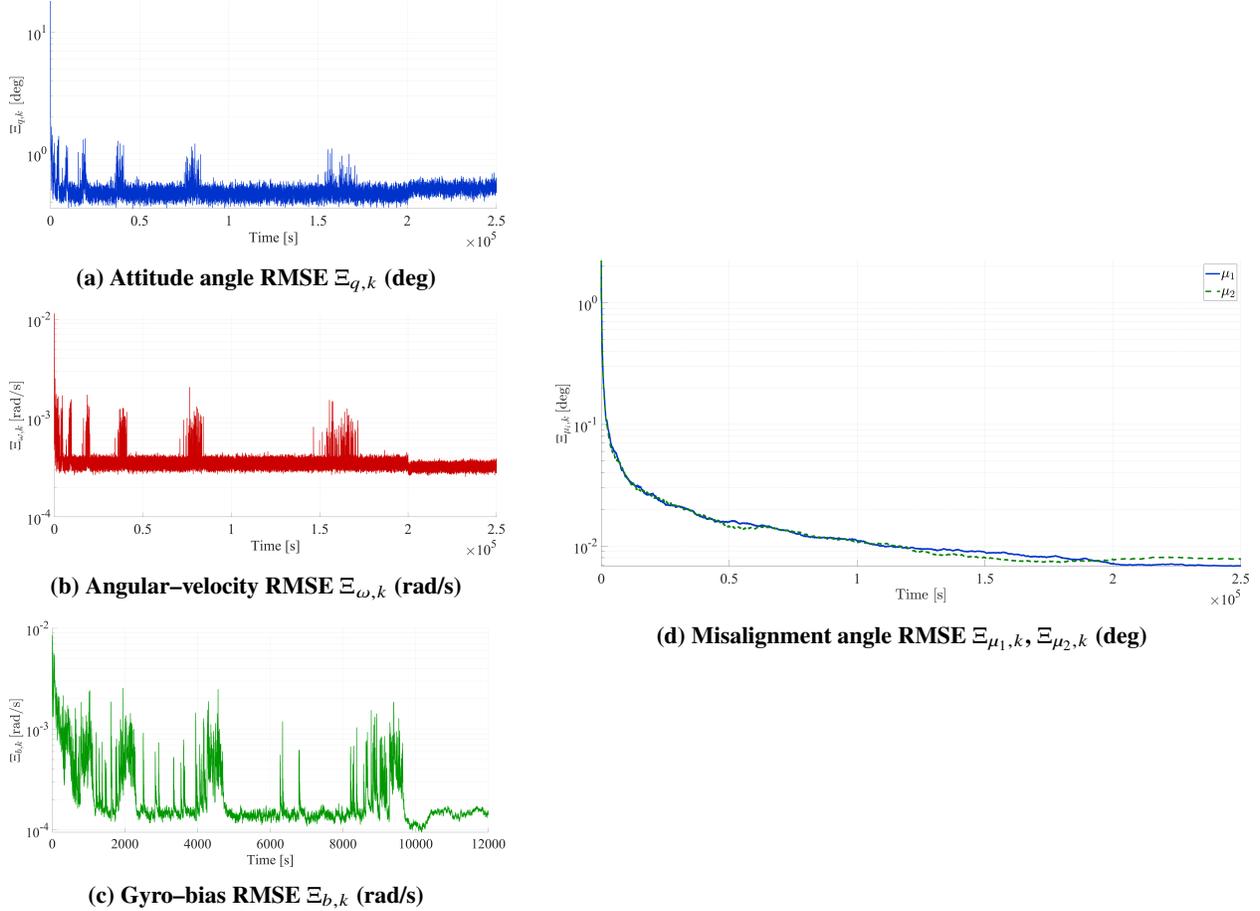

  \centering
  \begin{minipage}[c]{0.4\textwidth}
    \subfloat[Attitude angle RMSE $\Xi_{q,k}$ (deg)]{%
      \includegraphics[width=\textwidth]{Figures/attitudeRMSE2mis.png}%
      \label{fig:rmse_dual_att}}\\[1.2ex]
    \subfloat[Angular–velocity RMSE $\Xi_{\omega,k}$ (rad/s)]{%
      \includegraphics[width=\textwidth]{Figures/avelocityRMSE2mis.png}%
      \label{fig:rmse_dual_omega}}\\[1.2ex]
    \subfloat[Gyro–bias RMSE $\Xi_{b,k}$ (rad/s)]{%
      \includegraphics[width=\textwidth]{Figures/gyrofinalRMSE2mis.png}%
      \label{fig:rmse_dual_bias}}
  \end{minipage}
  \hfill
  \begin{minipage}[c]{0.56\textwidth}
    \centering
    \subfloat[Misalignment angle RMSE
              $\Xi_{\mu_1,k}$, $\Xi_{\mu_2,k}$ (deg)]{%
      \includegraphics[width=\textwidth]{Figures/2misRMSE2mis.png}%
      \label{fig:rmse_dual_mu}}
  \end{minipage}

  \caption{Monte Carlo RMSE metrics for the dual-misalignment case with $\Psi$–based refinement at $10\%$ and weighted–mean re–centering.}
  \label{fig:rmse_dual}
\end{figure}

Using Eq. \eqref{eq:rmse_q}–\eqref{eq:rmse_mu}, Fig.~\ref{fig:rmse_dual} reports the Monte Carlo RMSE trajectories for the dual–misalignment
case with $\Psi$–based refinement $(\Psi_\text{th}=10\%)$ and weighted–mean recentering of the model grid over the full time horizon.
Figure ~\ref{fig:rmse_dual_att} shows that the attitude angle RMSE $\Xi_{q,k}$ starts in the $\mathcal{O}(10)$~deg range because the initial attitude and misalignment hypotheses are only weakly informed. As the MEKF-MMAE locks onto the correct region of the grid during the first few calibration slews, $\Xi_{q,k}$ drops by more than an order of magnitude and, after this transient, remains well below 0.05 degrees for the rest of the run, with only intermittent spikes associated with the subsequent grid refinements. Indeed, after the MMAE is complete and convergence is reached, no spikes will be present in the estimation process. Panel~\ref{fig:rmse_dual_omega} reports the angular–velocity RMSE $\Xi_{\omega,k}$ in rad/s. The estimator quickly reduces the initial rate error from roughly $10^{-2}\,\mathrm{rad/s}$ to a baseline in the low $10^{-4}$–$10^{-3}\,\mathrm{rad/s}$ range and maintains this level for most of the time interval, with brief excursions toward $10^{-3}\,\mathrm{rad/s}$ whenever the dynamics are strongly excited due to the refinements. Panel~\ref{fig:rmse_dual_bias} shows a similar trend for the gyro–bias RMSE $\Xi_{b,k}$: the bias estimates converge steadily toward the true values, with a post–transient floor near $\mathcal{O}(10^{-4})\,\mathrm{rad/s}$ and short-lived peaks during the refinements.

Finally, Fig. ~\ref{fig:rmse_dual_mu} plots the misalignment angle RMSE for each star tracker, $\Xi_{\mu_1,k}$ and $\Xi_{\mu_2,k}$. Both curves start at about $2^\circ$ and decay smoothly by nearly four orders of magnitude, reaching more than a few $10^{-4}$~deg level by the end of the calibration sequence. The two traces
remain closely aligned over time, confirming that the joint 6D MMAE layer provides essentially symmetric calibration performance for both cameras under the chosen refinement policy. After the final braking–torque maneuver, the $\bm \mu_1$ curve settles slightly below $\mu_2$, which is consistent with the fact that the
$\mathrm{ST}_1$ triplet spans a wider angular separation on the celestial sphere. A wider, more nearly orthogonal star triad increases sensitivity to small
misalignments, leading to lower steady–state estimation error, in agreement with standard star–selection results for attitude determination ~\cite{MarkleyCrassidis2014},\cite{10.1117/12.807456}. In the Appendix, the full Monte Carlo consistency check for each state and parameter is reported in a double logarithmic scale, highlighting the robustness and accuracy of the MEKF-MMAE filter. It is worth mentioning that the convergence of the MEKF-MMAE filters is so accurate on the misalignment parameters that it enables the switch to those variables as additional states with a low error covariance, such that a new augment MEKF filter can run estimating the misalignment without the need for the MMAE bank of models after calibration is complete. 

The numerical values in Table~\ref{tab:final_mean_errors_dual} are consistent with these RMSE trends. The componentwise mean attitude errors are all below $0.06^\circ$ (on the order of $10^{-2}$~deg), while the mean misalignment errors for both cameras are below $6\times 10^{-4}$~deg in each axis (on the order of $10^{-4}$~deg). Likewise, the mean angular–velocity errors are on the order of $10^{-5}\,\mathrm{rad/s}$ and the mean gyro–bias errors are at or below
$10^{-5}\,\mathrm{rad/s}$, which is well within the steady–state bands observed in Fig.~\ref{fig:rmse_dual}. Together, these results indicate that the dual–tracker MMAE–MEKF achieves high–accuracy calibration of both misalignments without sacrificing attitude, rate, or bias estimation performance.

\begin{table}[tbh]
\centering
\caption{Final mean estimation errors averaged over the MC runs for the dual–misalignment case. Attitude and misalignment entries are reported as equivalent rotation angles derived from the MRP components; angular–rate and bias entries are componentwise errors in rad/s.}
\label{tab:final_mean_errors_dual}
\begin{tabular}{lccc}
\toprule
Quantity & $x$ & $y$ & $z$ \\
\midrule
\multicolumn{4}{c}{Angles (equivalent rotation, deg)} \\
\midrule
Attitude $\mathbf{q}$ [deg]
  & $-0.020942$ & $0.040991$ & $-0.054531$ \\
Misalignment $\boldsymbol{\mu}_1$ [deg]
  & $-0.000521$ & $0.000058$ & $-0.000012$ \\
Misalignment $\boldsymbol{\mu}_2$ [deg]
  & $-0.000512$ & $-0.000610$ & $0.000436$ \\
\midrule
\multicolumn{4}{c}{Angular–rate and bias components (rad/s)} \\
\midrule
Angular velocity $\boldsymbol{\omega}$ [rad/s]
  & $-2.835\times 10^{-5}$ & $-8.076\times 10^{-6}$ & $-7.210\times 10^{-6}$ \\
Gyro bias $\mathbf{b}$ [rad/s]
  & $2.735\times 10^{-6}$ & $-7.177\times 10^{-6}$ & $8.895\times 10^{-6}$ \\
\bottomrule
\end{tabular}
\end{table}

\section{Conclusion}
This paper presented a comprehensive pipeline and a novel adaptive multi-model framework for jointly estimating spacecraft attitude and star-tracker misalignments, tailored to GPS-denied, deep-space CubeSat missions. The approach combines a left-multiplicative MEKF with a Bayesian MMAE layer defined over a grid of body-to-sensor misalignment hypotheses, operating on TRIAD-based attitude observations in the single-misalignment case and on stacked line-of-sight measurements from two cameras in the dual-misalignment case. A consistent derivation of the error dynamics, measurement Jacobians, star-selection criteria, and a block-lower-triangular discretization enables misalignment compensation without augmenting the MEKF state with additional calibration parameters. A key algorithmic feature is the hypothesis-diversity metric $\Psi$, coupled with weighted-mean grid centering, which together enable adaptive refinement of the misalignment grid without premature convergence and substantially reduce final misalignment RMSE compared to classical MAP-based triggers, while remaining tractable for CubeSat-class onboard processors. The numerical results validate the proposed method and assess the filter as a valuable calibration technique. Monte Carlo simulations show that the resulting MEKF--MMAE architecture achieves arcsecond-level misalignment accuracy in the TRIAD-based single-misalignment configuration and recovers two independent misalignments with mean errors on the order of $0.4^{\prime\prime}$ per axis in the dual-misalignment configuration, while maintaining attitude errors below $180^{\prime\prime}$ and statistically consistent \mbox{$\pm 3\sigma$} performance for attitude, rate, bias, and both misalignments. Future work will extend the framework to slowly time-varying thermal or mechanical misalignment drifts, star-identification errors and dropouts, and hardware-in-the-loop demonstrations for operational deep-space attitude determination.

\section*{Acknowledgments}
The authors wish to acknowledge the support of this work through the National Aeronautics and Space Administration (NASA) Established Program to Stimulate Competitive Research (EPSCoR) under grant number 80NSSC24M0110.

\section{Appendix}
\begin{figure*}[tbh]
  \centering
  \begin{minipage}[t]{0.45\textwidth}
    \centering
    \subfloat[Attitude consistency]{%
      \includegraphics[width=\linewidth]{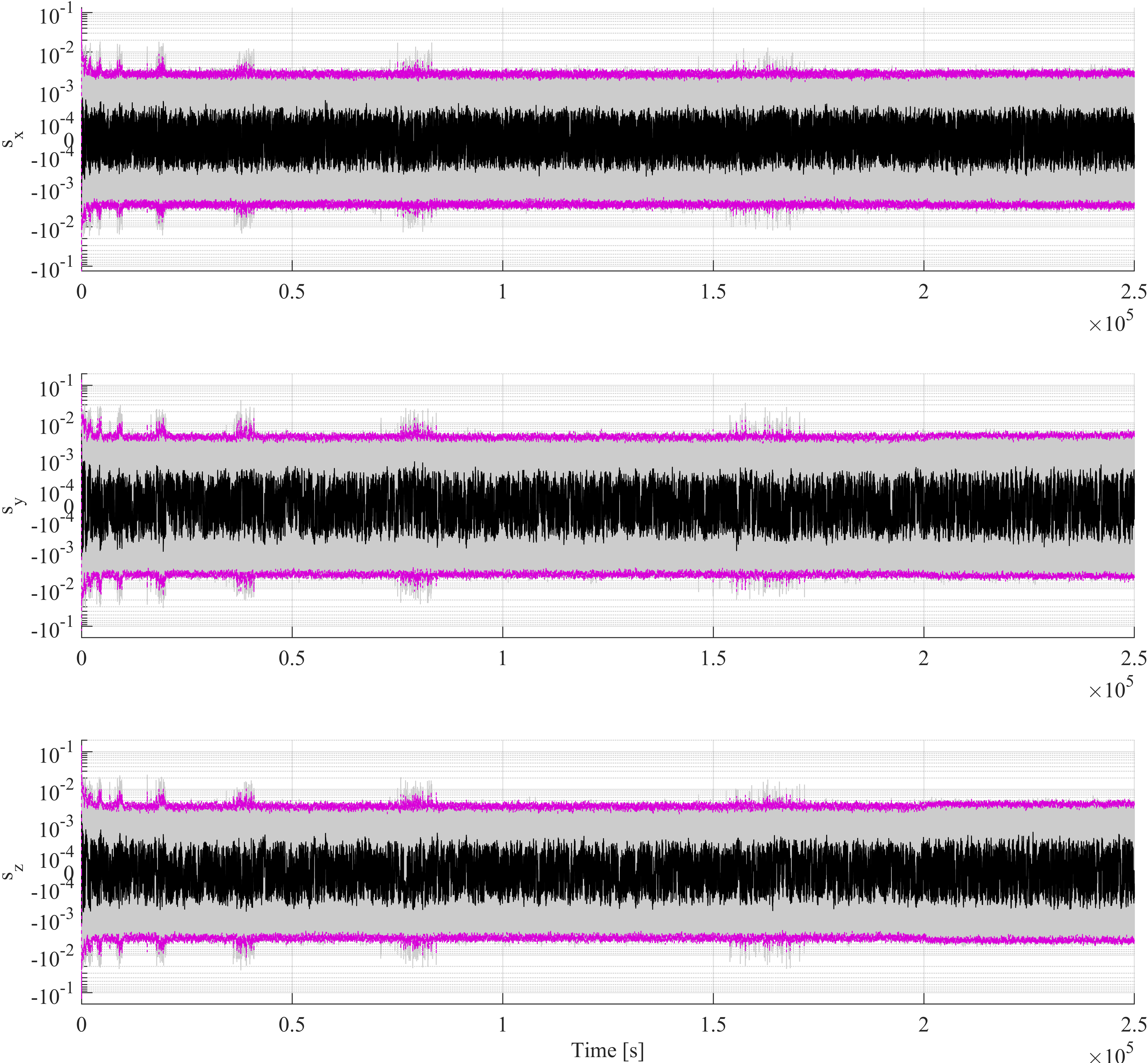}%
      \label{fig:cons_att_2mis}}\\[1.2ex]
    \subfloat[(Angular–velocity consistency]{%
      \includegraphics[width=\linewidth]{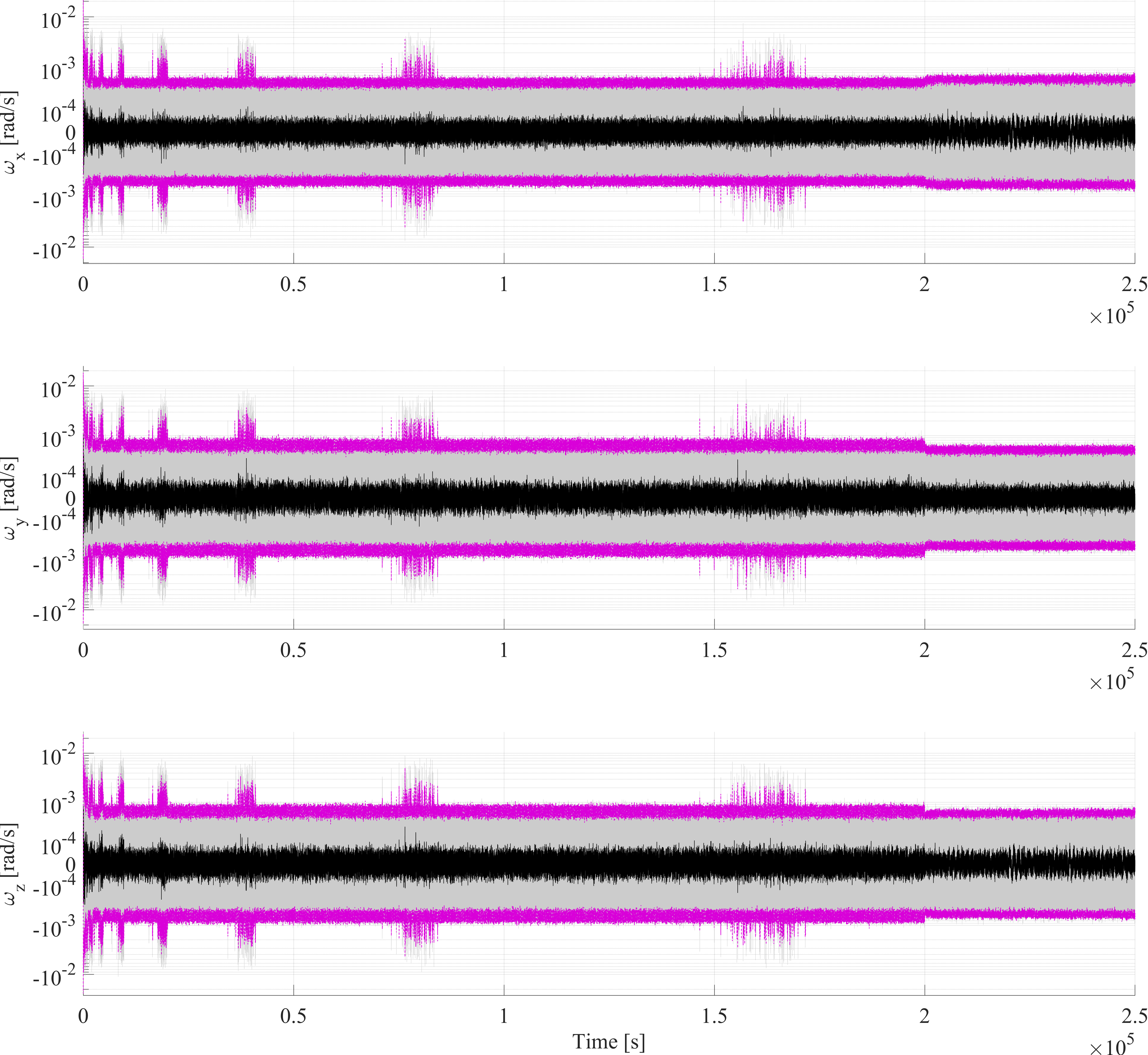}%
      \label{fig:cons_av_2mis}}\\[1.2ex]
    \subfloat[Gyro–bias consistency]{%
      \includegraphics[width=\linewidth]{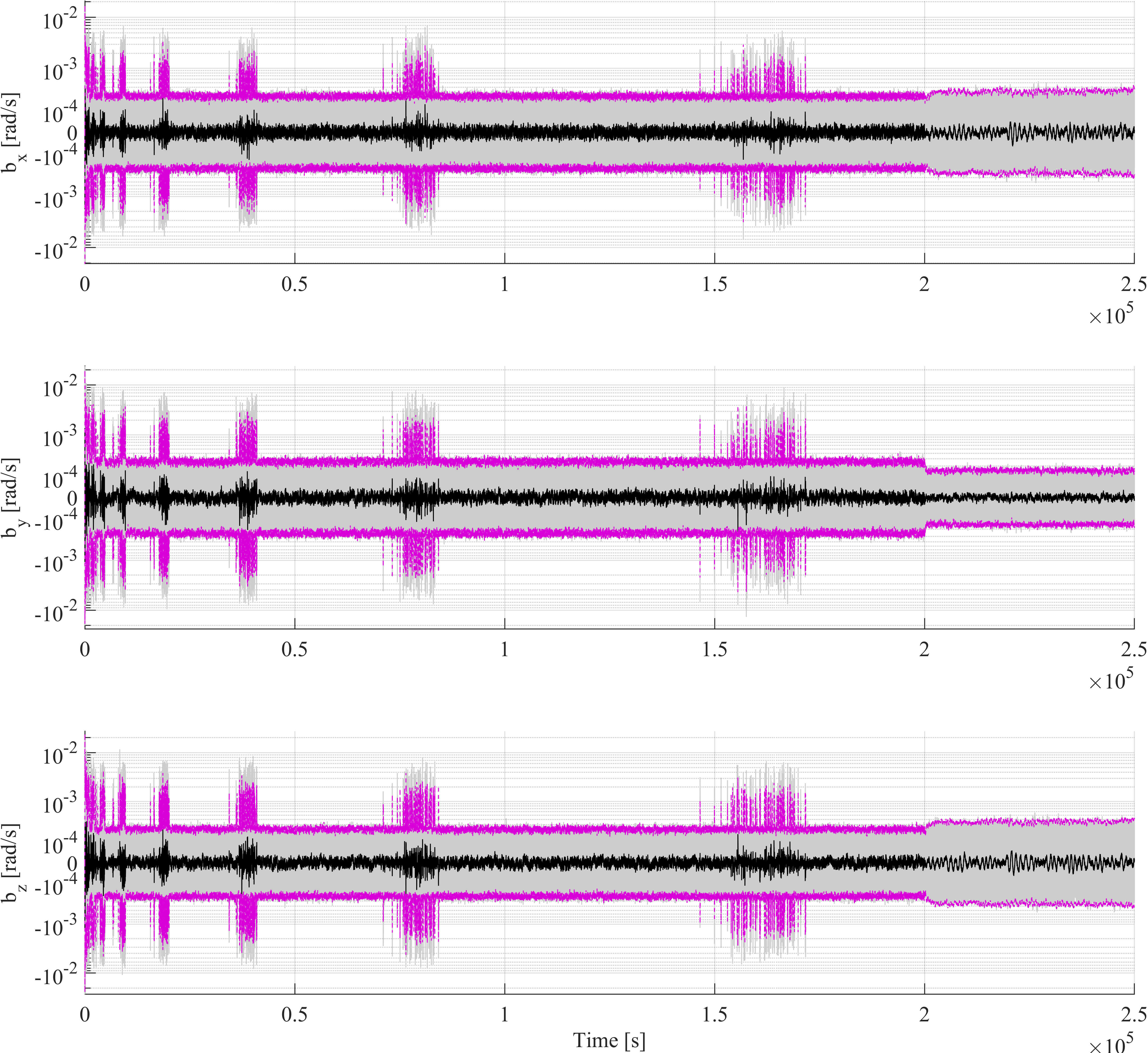}%
      \label{fig:cons_gyro_2mis}}
  \end{minipage}
  \hfill
  \begin{minipage}[t]{0.45\textwidth}
    \centering
    \vspace*{-0.05\textheight}
    \subfloat[Misalignment~1 consistency]{%
      \includegraphics[width=\linewidth]{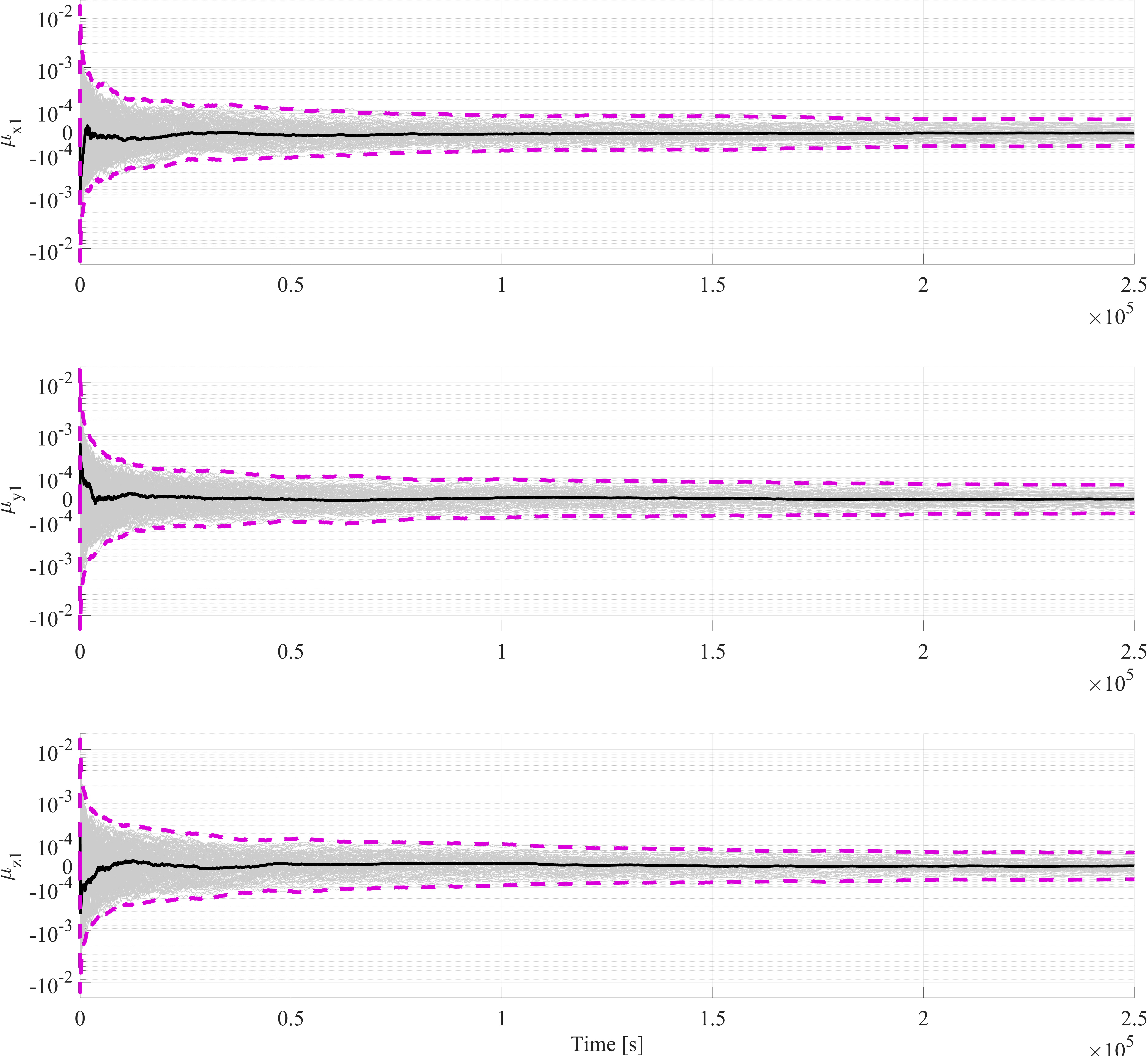}%
      \label{fig:cons_mu1_2mis}}\\[1.2ex]
    \subfloat[Misalignment~2 consistency]{%
      \includegraphics[width=\linewidth]{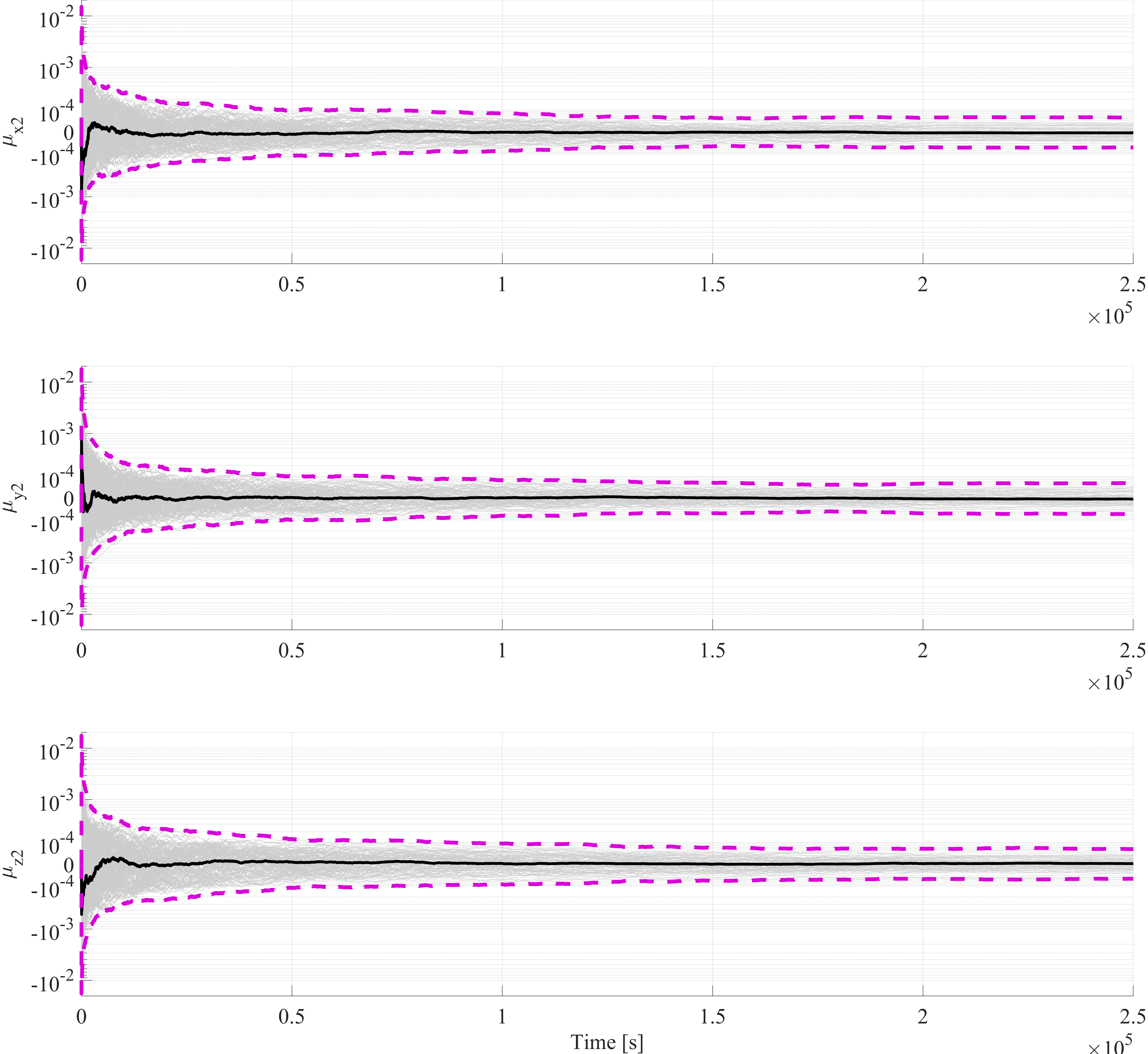}%
      \label{fig:cons_mu2_2mis}}
  \end{minipage}

  \caption{Consistency analysis for the dual–misalignment MMAE–MEKF case:
  (a) attitude error, (b) angular–velocity error, (c) gyro–bias error,
  and (d–e) misalignment errors for star trackers~1 and~2, respectively.}
  \label{fig:consistency_dual}
\end{figure*}

The consistency and accuracy of the MEKF-MMAE filter are assessed by analyzing the statistical behavior of the filter, examining the estimation error for each component
\begin{equation}
  e_i(t) \triangleq \hat{x}_i(t) - x_i^{\text{true}}(t),
  \label{eq:component_error}
\end{equation}
across all the Monte Carlo runs. For each state component $i$ and time sample $t_k$, we compute the sample mean $\bar e_i(t_k)$ and standard deviation $\sigma_i(t_k)$ over the ensemble, and Fig.~\ref{fig:consistency_dual} visualizes this behaviour. The gray traces show the individual estimation errors $e_i^{(\ell)}(t)$ for each
run $\ell$, the solid black curve is the Monte Carlo mean $\bar e_i(t)$, and the magenta dashed curves denote the $\bar e_i(t) \pm 3\,\sigma_i(t)$ envelopes. A bi–symmetric logarithmic (``symlog'') scaling (\cite{webber},\cite{doi:10.2514/1.A35688}) is used on the vertical axes to resolve both small deviations near zero and rare larger excursions while preserving the sign of the error.

In the left column (attitude MRPs, angular velocity, and gyro bias), the estimation errors exhibit a short initial transient associated with the large prior uncertainty and early calibration slews (roughly the first $1$–$2\times10^{4}\,\text{s}$), after which the mean remains tightly clustered around zero and almost all trajectories stay inside the
$\pm 3\sigma$ bounds. Occasional spikes during the refinements are there and approximately symmetric in sign, as expected for this problem and a well–tuned stochastic model. The right column shows the same test for the two misalignment vectors $\boldsymbol{\mu}_1$ and $\boldsymbol{\mu}_2$. After the early transient, all components rapidly contract from $\mathcal{O}(10^{-1})$ at $t=0$ into the $10^{-5}$–$10^{-6}$ range and remain symmetrically distributed within the $\pm 3\sigma$ envelopes for the remainder of the simulation. The absence of drift or systematic offset in these errors indicates that the MMAE bank weighting, refinement strategy, and MEKF covariance propagation provide an accurate uncertainty description for both the fast dynamical states and the slowly varying calibration parameters. Together with the RMSE trends in Fig.~\ref{fig:rmse_dual}, this demonstrates that the dual–misalignment MMAE–MEKF is both accurate and statistically consistent over the full simulation horizon. 

\bibliography{sample}

\end{document}